%% file: main.tex
\documentclass[sigconf]{acmart}
\AtBeginDocument{%
  \providecommand\BibTeX{{%
    \normalfont B\kern-0.5em{\scshape i\kern-0.25em b}\kern-0.8em\TeX}}}

\setcopyright{acmcopyright}
\copyrightyear{2019}
\acmYear{2019}
\acmDOI{10.1145/1122445.1122456}

\acmConference[Under Publication Review]{}{June 20}{2019}
\acmConference[CCS '19]{CCS '19: ACM Conference on Computer and Communications Security}{November 11--15, 2019}{London, UK}
\acmBooktitle{CCS '19: ACM Conference on Computer and Communications Security, November 11--15, 2019, London, UK}
\acmPrice{15.00}
\acmISBN{978-1-4503-9999-9/18/06}

\acmConference[Under Publication Review]{}{June 20}{2019}
\acmPrice{}
\acmISBN{}

\setcopyright{none}
\renewcommand\footnotetextcopyrightpermission[1]{}
\usepackage{capt-of}
\usepackage{color, colortbl}
\usepackage{pifont}
\usepackage{makecell}
\usepackage{tikz}
\graphicspath{{figures/}}

\newcommand{\figline}{\vspace{3pt}\hrulefill}

\definecolor{GreenHLight}{RGB}{21,176,79}
\definecolor{RedHLight}{RGB}{250,34,0}
\definecolor{BlueHLight}{RGB}{53,176,239}
\definecolor{YellowHLight}{RGB}{251,217,102}

\newcommand{\cmark}{\ding{51}}%
\newcommand{\xmark}{\ding{55}}%

\newcommand{\icad}{ICAS}
\newcommand{\icadplural}{ICAS'}
\newcommand{\icadpartiallyexpanded}{IC Attack Surface}

\begin{document}
\title{An Extensible Framework for Quantifying the Coverage of Defenses Against Untrusted Foundries}

\author{Timothy Trippel}
\authornote{Work done at MIT Lincoln Laboratory.}
\author{Kang G. Shin}
\email{{trippel,kgshin}@umich.edu}
\affiliation{
    \institution{University of Michigan}
    \city{Ann Arbor}
    \state{Michigan}
}

\author{Kevin B. Bush}
\email{kevin.bush@ll.mit.edu}
\affiliation{
    \institution{MIT Lincoln Laboratory}
    \city{Lexington}
    \state{Massachusetts}
}

\author{Matthew Hicks}
\authornotemark[1]
\authornote{Corresponding faculty author}
\email{mdhicks2@vt.edu}
\affiliation{
    \institution{Virginia Tech}
    \city{Blacksburg}
    \state{Virginia}
}

\begin{abstract}
\input{sections/0_abstract}
\end{abstract}



\keywords{Hardware Security; Fabrication-time Attacks and Defenses; VLSI}

\maketitle

\section{Introduction}\label{section:introduction}
\input{sections/1_introduction}

\section{Background}\label{section:background}
\input{sections/2_background}

\section{Threat Model}\label{section:threat_model}
\input{sections/3_threat_model}

\input{sections/3_5_defenseOverview}

\section{Unified Attack Metrics}\label{section:design}
\input{sections/4_design}

\section{An Extensible Coverage Assessment Framework}\label{section:implementation}
\input{sections/5_implementation}

\section{Evaluation}\label{section:evaluation}
\input{sections/6_evaluation}

\section{Discussion}\label{section:discussion}
\input{sections/7_discussion}

\section{Related Work}\label{section:related_work}
\input{sections/8_related_work}

\section{Conclusion}\label{section:conclusion}
\input{sections/9_conclusion}

\section*{Acknowledgments}
\input{sections/10_acknowledgement}

\bibliographystyle{ACM-Reference-Format}
\bibliography{sections/bibliography.bib}

\clearpage
\appendix
\section{Route Distances of OR1200 Layouts}\label{section:appendix_1}
\input{sections/11_appendix_1}

\end{document}

%% file: sections/0_abstract.tex
The transistors used to construct Integrated Circuits (ICs) continue to shrink. While this shrinkage improves performance and density, it also reduces trust: the price to build leading-edge fabrication facilities has skyrocketed, forcing even nation states to outsource the fabrication of high-performance ICs. Outsourcing fabrication presents a security threat because the black-box nature of a fabricated IC makes comprehensive inspection infeasible. Since prior work shows the feasibility of fabrication-time attackers' evasion of existing post-fabrication defenses, IC designers must be able to protect their physical designs before handing them off to an untrusted foundry. To this end, recent work suggests methods to harden IC layouts against attack. Unfortunately, no tool exists to assess the effectiveness of the proposed defenses---meaning gaps may exist.

This paper presents an extensible IC layout security analysis tool called {\em \icadpartiallyexpanded{}} (\icad{}) that quantifies defensive coverage. For researchers, \icad{} identifies gaps for future defenses to target, and enables the quantitative comparison of existing and future defenses. For practitioners, \icad{} enables the exploration of the impact of design decisions on an IC's resilience to fabrication-time attack. \icad{} takes a set of metrics that encode the challenge of inserting a hardware Trojan into an IC layout, a set of attacks that the defender cares about, and a completed IC layout and reports the number of ways an attacker can add each attack to the design. While the ideal score is zero, practically, our experience is that lower scores correlate with increased attacker effort.

To demonstrate \icadplural{} ability to reveal defensive gaps, we analyze over 60 layouts of three real-world hardware designs (a processor and AES and DSP accelerators), protected with existing defenses. We evaluate the effectiveness of each circuit/defense combination against three attacks from the literature. Results show that some defenses are ineffective and others, while effective at reducing the attack surface, leave 10's to 1000's of unique attack implementations for an attacker to exploit.

%% file: sections/1_introduction.tex
The relationship between complexity and security seen in software also holds for Integrated Circuits (ICs). Since the inception of the IC, transistor sizes have continued to shrink. For example, compare the 10\,${\mu}m$ feature size of the original Intel 4004 processor~\cite{intel_processor_history} to the 10\,$nm$ feature size of Intel's recently announced Ice Lake processor family~\cite{ice_lake_by_june}. Smaller transistors enable IC designers to create increasingly complex circuits with higher performance and lower power-usage. However, continuing this trend pushes the laws of physics and comes at a substantial cost: by 2020, the cost to build a leading-edge fabrication facility is estimated to be \$15--20B~\cite{cost_of_fab_2020}.

Such costs are prohibitive for not only most semiconductor companies, but also nation states. Thus, \textbf{most hardware design houses are fabless}, i.e., while they are able to fully design and lay out an IC, they must outsource its fabrication. Outsourcing combined with the black-box nature of a fabricated IC requires fabless semiconductor companies to trust that their physical designs will not be altered maliciously by the foundry, also known as a \textit{fabrication-time attack}. Previous work demonstrates several ways a fabrication-time attacker can insert a hardware Trojan into an otherwise trusted IC~\cite{becker2013stealthy,kumar2014parametric,a2}. A2~\cite{a2} demonstrates the most stealthy and controllable IC fabrication-time attack to date, whereby a hardware Trojan with a complex, yet stealthy, analog trigger circuit is inserted into the finalized layout of a processor. Even though the inserted Trojan is small, the attacker can trigger it and escalate to a persistent software-level attack (i.e., a hardware foothold~\cite{king08}) using only user-mode code.

Early work focuses on post-fabrication \emph{detection} of hardware Trojans in ICs~\cite{tehranipoor2010survey}. Broadly, there are two classes of detection: 1) side-channel analysis and 2) Trojan-activation via functional testing. Side-channel (power, timing, etc.) analysis~\cite{agrawal2007trojan,jin2008hardware,potkonjak2009hardware,narasimhan2011tesr} assumes that the Trojan's trigger is complex (i.e., many logic gates); thus, noticeably changes the physical characteristics of the chip. For example, inserting the large amount of extra logic required by a complex trigger into a design alters the power signature of the device. Alternatively, Trojan-activation via functional testing assumes that the Trojan's trigger is simple (i.e., few logic gates~\cite{becker2013stealthy,kumar2014parametric}); thus, easily activated by test vectors. Unfortunately, layering detection classes is \emph{not} sufficient as recent work shows that it is possible to create an attack that is both small and stealthy~\cite{a2}.

To address the gaps left by post-fabrication Trojan \emph{detection} schemes, recent work focuses on pre-fabrication, IC layout-level, Trojan \emph{prevention}~\cite{ba2016hardware,ba2015hardware,cocchi2014circuit,xiao2013bisa}. IC layout-level defenses work by: 
\begin{enumerate}
    \item increasing placement \& routing resource utilization
    \item increasing congestion around security-critical design components
\end{enumerate}
The lack of resources deprives the attacker of the required transistors needed to implement their Trojan trigger/attack circuits, and the increased congestion around security-critical wires acts as a barrier for the attacker attempting to integrate their Trojan into the victim design. Ideally, defenders utilize just enough resources and create enough congestion such that the attacker cannot implement and insert their attack, while keeping the design routable. Short of that, the added barriers require the attacker to expend more resources (e.g., time) to insert their attack into an IC layout.\footnote{Time is the most critical resource for the attacker as IC fabrication is bounded in terms of turnaround time.}

Two IC layout-level defensive approaches exist: 1) undirected and 2) directed. \textbf{Undirected} approaches aim to (probabilistically) increase resource utilization and congestion across the \textit{entire} layout by altering existing place-and-route parameters (e.g., core density~\cite{xiao2013bisa}) that will likely result in increased resource utilization and congestion. More recently, a line of \textbf{directed} approaches have emerged~\cite{ba2015hardware,ba2016hardware} that \textit{systematically} increase utilization of \textit{specific-regions} of the device layer, i.e., nearby security-critical components. Given that it is infeasible to occupy the entire device layer in a tamper-evident manner~\cite{ba2016hardware} both classes of approaches may leave IC layouts vulnerable to attack by an untrusted foundry.

To identify gaps in existing defenses and guide future IC layout-level defenses, we design and implement an extensible measurement framework that quantifies defensive coverage with respect to specific foundry-level attacks. Our framework, {\em \icadpartiallyexpanded{}} (\icad{}), quantifies defensive coverage in three dimensions that capture the essence and difficulty of inserting a hardware Trojan at an untrusted foundry:
\begin{enumerate}
\item \textbf{Trojan logic placement:} finding unused space to place additional circuit components
\item \textbf{Victim/Trojan integration:} attaching hardware Trojan payload to security-critical logic
\item \textbf{Intra-Trojan routing:} connecting the trigger and payload portions of the hardware Trojan
\end{enumerate}
A successful attack requires all three steps.

Using \icad{}, we analyze over 60 different IC layouts across three fully-functional ASIC designs: an AES accelerator, a DSP accelerator, and an OR1200 processor. For each layout, \icad{} reports the coverage against four attacks~\cite{a2,hicks10,king08,trusthub} that span the digital and analog domain as well a range of attack outcomes. \icad{}'s analysis reveals that all existing IC layout-level defenses are incomplete, leaving 1000's of opportunities for an attacker at an untrusted foundry to insert a hardware Trojan. An additional finding is that even though most existing countermeasures do increase the complexity of inserting a hardware Trojan, some countermeasures are ineffective. Lastly, \icad{}'s analysis suggests that focusing on exhausting resources on the device layer (i.e., transistors) is an incomplete defense; future defenses should also aim to increase congestion around security-critical wires.

This paper makes the following contributions:
\begin{itemize}
    \item We propose an extensible methodology that quantifies the difficulty of inserting hardware Trojans into an existing IC layout by an untrusted foundry.
    \item We design, implement, and open-source~\cite{anon} our extensible framework, \icad{}, that computes various layout-specific security metrics. The \icad{} framework provides an interface to programmatically query the physical layout of an IC (encoded in the GDSII format) to compute various security metrics with respect to attacks-of-interest.
    \item We use \icad{} to quantify the effectiveness and expose the defensive gaps of previously-proposed untrusted foundry defenses by analyzing over 60 IC layouts of three real-world hardware cores.
    \item We identify future directions for defenses that work in a layered fashion with existing defenses.
\end{itemize}

%% file: sections/2_background.tex
\begin{figure}[t]
\centering
\includegraphics[width=0.25\textwidth]{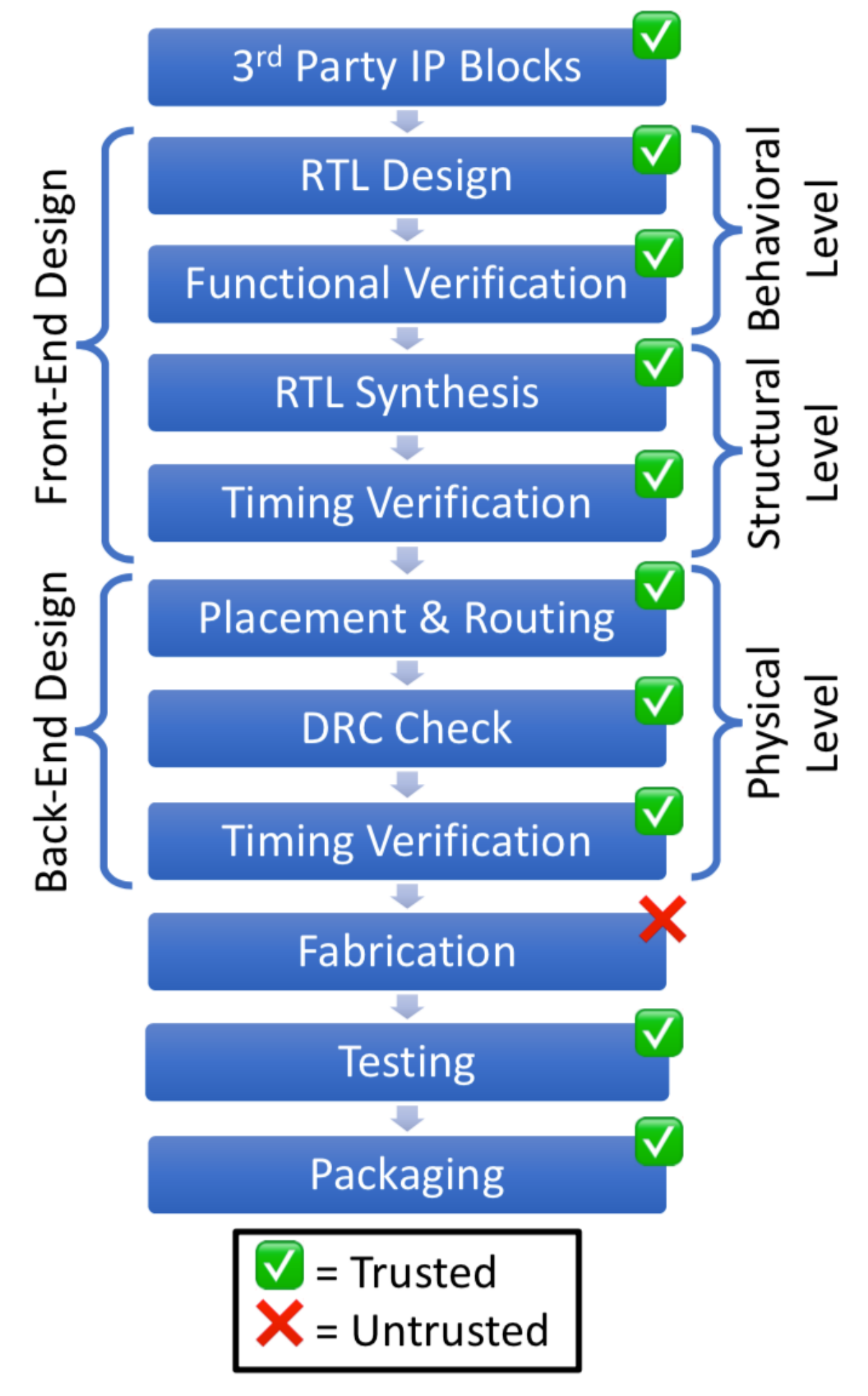}
\vspace*{-0.05in}
\caption{\footnotesize The typical IC design process starts with a textual specification of design requirements and ends with a fabricated and tested chip. Green check-boxes mark trusted stages and red x-boxes mark the untrusted step (i.e., an untrusted foundry). The fabrication step takes a GDSII file (physical IC layout) as input and produces a wafer of die. While prior work proposes metrics for untrusted front-end design~\cite{hicks10,waksman2013fanci,salmani2013analyzing,chakraborty2009mero}, no mechanism exists for measuring an IC layout's resilience to an untrusted foundry.}
\label{fig:ic_design_flow}
\figline{}
\vspace*{-0.15in}
\end{figure}

\subsection{IC Design Process}
Figure~\ref{fig:ic_design_flow} shows the typical IC design
process~\cite{rostami2013hardware}, which consists of
three main phases: 1) front-end design, 2) back-end design, and 3) fabrication. The front-end design phase can be further split into two design abstraction levels, \textit{behavioral} and \textit{structural}, while a single design abstraction level, \textit{physical} (i.e., consists of both analog and digital properties), encompasses the back-end. The front-end design process begins by first describing the functionality of the circuit at the behavioral level, also known as the Register Transfer Level (RTL), using a hardware
description language (HDL), like VHDL or Verilog. Next, the behavioral level description of the circuit is transformed into a structural level description during RTL synthesis. RTL synthesis is similar to software compilation: the RTL design is optimized and reduced to a set of logically connected
digital logic gates, called a {\em gate-level netlist} (netlists are
commonly described using an HDL language). The gate-level netlist is then passed to the back-end design phase to be transformed into something able to be implemented into a physical chip (i.e., an IC layout) through a process known as Placement and Routing (PaR). 

\begin{figure}[t]
\centering
\includegraphics[width=0.32\textwidth]{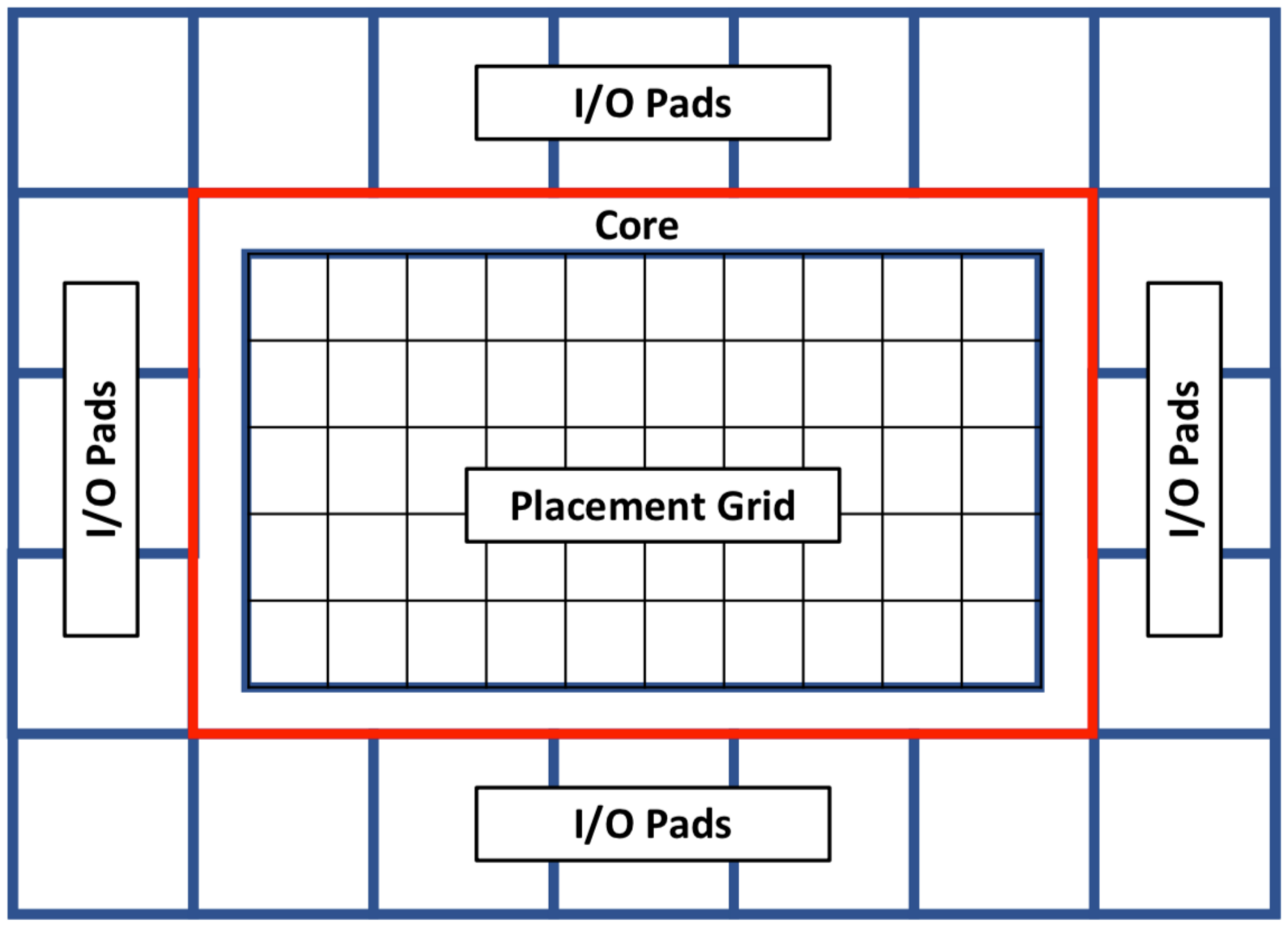}
\caption{\footnotesize Typical IC floorplan created during the place-and-route design phase. The floorplan consists of an I/O pad ring surrounding the chip core. Within the core is the placement grid. Circuit components are placed and routed within the placement grid.}
\label{fig:ic_floorplan}
\figline{}
\vspace*{-0.15in}
\end{figure}

IC layouts consist of multiple layers. The bottom layers are \textit{device layers}, while the top layers are \textit{metal layers}. Device layers are used for constructing circuit components (e.g., transistors), and the metal layers are used for routing (e.g., vias and wiring). The first stage of PaR is creating a floorplan. Figure~\ref{fig:ic_floorplan} illustrates an IC floorplan. To create a floorplan, the dimensions of the overall chip are specified and the core area is defined. Typically a ring of I/O pads is then placed around the chip core, while a placement grid is drawn over the core. Each tile in the placement grid is known as a \textit{placement site}. Circuit components (e.g., standard cells) are then placed on the placement grid, occupying one or more placement sites, depending on the size of the component. Lastly, all components are routed together, using one or more routing layers. The output from the back-end design is a Graphics Database System II (GDSII) file that is a geometric description of the placed-and-routed circuit layout. The GDSII file is then sent to a fabrication facility where it is manufactured. The final step is testing and packaging.

\subsection{Hardware Trojans}\label{section:hardware_trojans}
\subsubsection{\textbf{Trojan Components}}
\begin{figure}[t]
\centering
\includegraphics[width=0.48\textwidth]{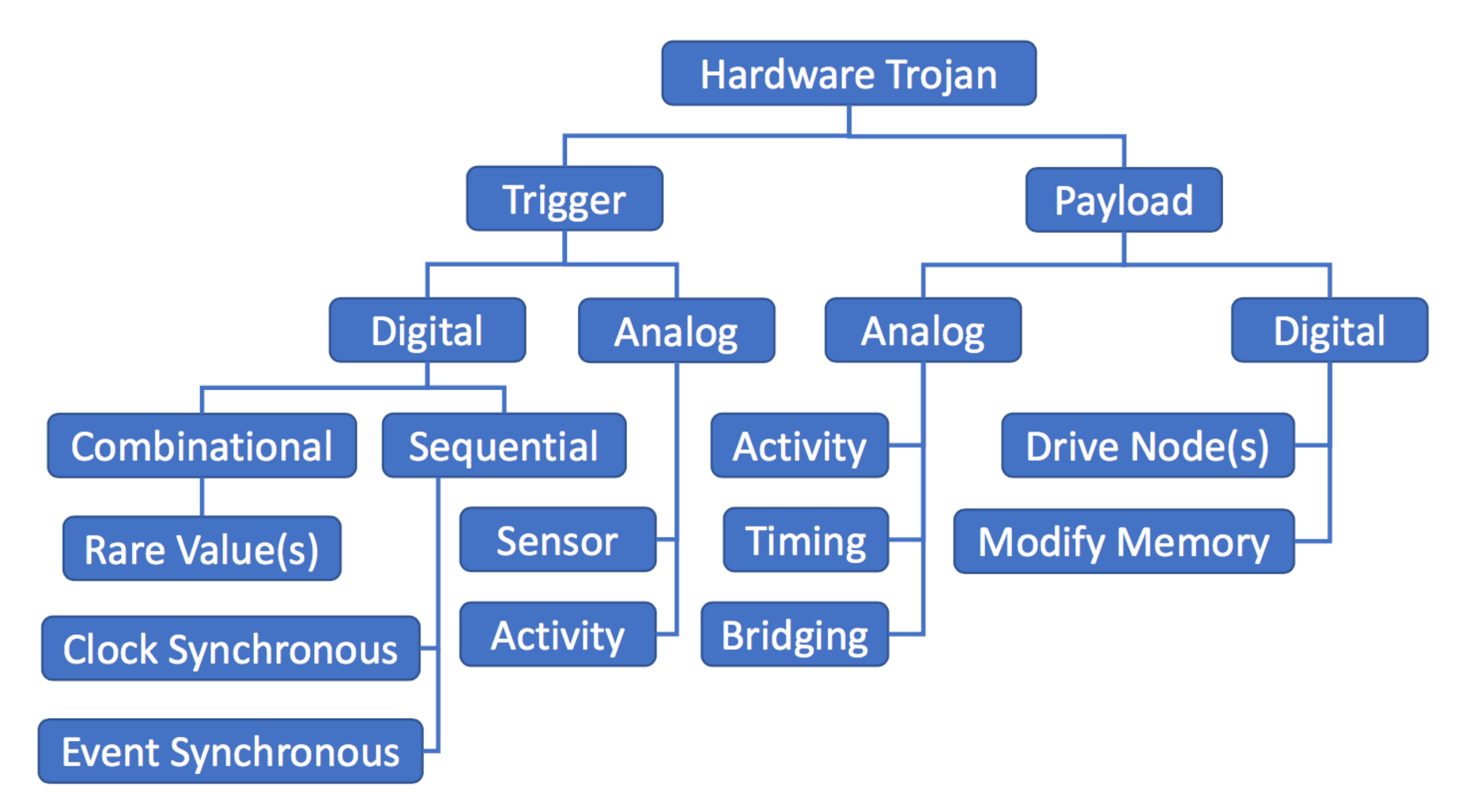}
\caption{\footnotesize An existing taxonomy
of hardware Trojans~\cite{chakraborty2009hardware}. This
taxonomy classifies hardware Trojans based on their \textit{trigger}
and \textit{payload} types.}
\label{fig:hw_trojan_taxonomy}
\figline{}
\vspace*{-0.15in}
\end{figure}

A hardware Trojan is a malicious modification to 
a circuit designed to modify its behavior during
operation~\cite{beaumont2011hardware}. Hardware Trojans have two main components: 1) \textbf{trigger} and 
2) \textbf{payload}~\cite{chakraborty2009hardware,jin2008hardware,wolff2008towards}. Prior
work classifies hardware Trojans based on the
functionalities of their trigger and payload mechanisms~\cite{chakraborty2009hardware,jin2008hardware,wolff2008towards}. 
In this paper, we adopt and simplify an existing hardware Trojan taxonomy~\cite{chakraborty2009hardware}; shown in Figure~\ref{fig:hw_trojan_taxonomy}.

The \textbf{trigger} mechanism of a hardware Trojan is what
\textit{initiates} the delivery of the Trojan's payload. Triggers
can be built by adding, removing, or altering existing
hardware in an IC. They can be digital~\cite{king08} or analog~\cite{a2}. The ideal trigger is \textit{small}: requiring few or no additional circuit components, \textit{stealthy}:
requiring dozens of rare events to activate, and \textit{controllable}: readily attacker deployable, but not so by defenders or through regular use. There have been several
triggers demonstrated before that span the trade-space of large
(requiring many additional gates) and stealthy~\cite{lin2009trojan} to the
opposite: small (requiring no additional gates) and easy to
trigger~\cite{shiyanovskii2010process,becker2013stealthy}. The most advanced Trojans are small, stealthy, and controllable~\cite{a2}.

The \textbf{payload} mechanism receives a signal from the
trigger and alters the functionality of the IC. Analog~\cite{shiyanovskii2010process,becker2013stealthy} and digital~\cite{a2} payloads exist, with a variety of effects. These effects can
leak information~\cite{lin2009trojan}, alter the internal state of the IC~\cite{a2},
or cause a system to be unusable (denial-of-service)~\cite{shiyanovskii2010process}. Regardless of effect, the payload mechanism must route a wire to, or in the
vicinity of, some target ``security-critical''~\cite{specs15} wire in the IC design.

\subsubsection{\textbf{Trojan Implementations}}
There are three ways a malicious foundry can "insert" a hardware Trojan into an otherwise trusted IC layout: \emph{additive}, \emph{substitution}, and \emph{subtractive}. Additive Trojans involve inserting additional circuit components and/or wiring into an existing design. Substitution Trojans require removing logic with low observability to make room for additional Trojan circuit components and/or wiring in an existing circuit design. Lastly, subtractive Trojans require removing circuit components and/or wiring to alter the behavior of a existing circuit design. The focus of this paper is \textbf{assessing defensive coverage with respect to additive Trojans}. Substitution and subtractive Trojans, while intriguing, remain largely unexplored by the community. We do not know of any demonstrably stealthy and controllable substitution or subtractive Trojans and when researchers do create such an attack, there exists orthogonal mitigation strategies~\cite{wang2008detecting}.\footnote{Dopant-level Trojans are the closest substitution Trojan designs demonstrated in the literature~\cite{becker2013stealthy,kumar2014parametric}. Though their non-existent footprints make them difficult to detect via side channels, their simplistic designs and limited controllability make them detectable during post-fabrication testing~\cite{sugawara2014reversing}.}

Inserting an additive Trojan at an untrusted foundry requires modifying two fundamental characteristics of an IC's physical layout---placement and routing---regardless of how a an attacker implements the Trojan's trigger and payload.
We define \textbf{\textit{Trojan placement}} to be the act of placing additional hardware components into an IC layout for the purpose of crafting a Trojan trigger and payload, \textbf{\textit{Victim/Trojan integration}} to be wiring the Trojan's payload to, or in the vicinity, of a security-critical net in the victim IC layout, and \textbf{\textit{intra-Trojan routing}} to be the act of wiring the hardware Trojan together.
The most challenging aspect of inserting a hardware Trojan at fabrication-time is finding empty space on the IC's device layer to insert the trigger and payload components (\textbf{Trojan placement}), \underline{AND} routing the payload to a security-critical net (\textbf{Victim/Trojan integration}). \icad{} quantifies each of these fundamental tasks, in turn identifying weak points in the IC layout that an attacker might exploit.

%% file: sections/3_threat_model.tex
We adopt a threat model for untrusted foundry attacks that assumes all steps in the IC design process can be trusted, \textit{except} for all of the processes---no matter if they are outsourced---performed by a foundry (colloquially, fabrication). Figure~\ref{fig:ic_design_flow} depicts our threat model. This entails that the RTL is designed, synthesized, and laid-out by trusted parties. Post fabrication testing is also performed by a trusted party. We adopt this threat model since the astronomical costs to fabricate ICs force most semiconductor companies to outsource fabrication. To this point, in 2005, the U.S. government identified the untrusted foundry threat as the most significant weakness of the microelectronics supply chain~\cite{force2005high}. 

We restrict our threat model to fabrication-time attacks involving \emph{additive} Trojans, i.e., hardware Trojans that require inserting additional circuitry to a physical IC design. Previous work on substitution/subtractive hardware Trojans shows that such Trojan insertion methods are addressable by measuring the controllability and observability of logic at the \emph{behavioral} and/or \emph{structural} level of the IC design, for which several methods have already been proposed~\cite{salmani2017cotd,zhang2015veritrust,ccakir2015hardware,salmani2013analyzing,waksman2013fanci,hicks10,goldstein1980scoap}. Orthogonally, this work fills the void of quantifying the susceptibility of an IC design to additive hardware Trojan insertion at the \emph{physical} level of the IC design process by an untrusted foundry.

Focusing on additive hardware Trojans, an adversary can only insert additional components/wires. They cannot increase the size of the chip to make additional room for the implants because this is readily caught by defenders. As a result, an attacker has two choices: \textit{find} open space in the design large enough to accommodate the additional circuitry, or \textit{create} open space in the design by moving circuitry around. The latter is extremely challenging due to its recursive nature, it runs the risk of violating fragile timing constraints and manufacturing design rules, and it increases fabrication turnaround time (which is usually set to three months); any of which could expose the Trojan. Therefore, our focus is identifying open spaces suitable for hardware Trojan implementation.

%% file: sections/3_5_defenseOverview.tex
\section{Untrusted Foundry Defenses}

To protect IC layouts against insertion of a hardware Trojan by attackers at an untrusted foundry, two classes of defenses exist: \textbf{undirected} and \textbf{directed}.
Undirected defenses leverage existing tuning knobs available during the IC layout process, but do not differentiate between security-critical and general-purpose wires and logic.
Thus, undirected approaches provide probabilistic protection.
On the other hand, directed defenses require augmenting existing PaR tool flows to harden the resulting IC layout, focusing on deploying defenses systematically around security-critical wires and logic.
Thus directed approaches provide targeted protection, but increase the complexity of the place-and-route process.

This section provides an overview of the landscape of undirected and directed defenses.
The focus is the mechanism each defense uses to increase the complexity faced by a foundry-level attacker.
We use the results of the defensive analysis in this section to develop a set of unifying coverage metrics in the next section.
Finally, in the evaluation, we evaluate commercial IC layouts using the defense-inspired metrics to quantify each defense's coverage.

\subsection{Undirected}
\label{sec:undirected}

The lowest cost approach for protecting an IC layout from a foundry-level attacker is to take advantage of existing physical layout parameters (e.g., core density, clock frequency, and max transition time) offered by commercial CAD tools~\cite{xiao2013bisa,innovus}.
The goal is to increase congestion across the component layer and the routing layer.
Ideally, this also results in increased congestion around security-critical logic and wires.
Practically, increases in congestion around security-critical logic and wires is probabilistic.

Increased congestion is a symptom of increased resource utilization; hence, there are fewer resources available to the attacker.
The most obvious resource that an attacker cares about are placement sites on the component layer.
Increasing the density, \emph{decreases unused placement sites}.
Without sufficient placement sites, the attacker cannot implement their Trojan logic.
A less obvious resource is attachment points on security-critical wires that serve as victim/Trojan integration points..
Increasing routing layer congestion (via density and/or timing constraints) \emph{increases the blockage around security-critical wires}, meaning there are less integration points.

\subsection{Directed}
\label{sec:directed}

To address the shortcoming of undirected approaches, recent defenses advocate focusing on security-critical logic and wires.
Specifically, the approaches aim to prevent the attacker from being able to implement their hardware Trojan by occupying unused placement sites (i.e., transistors)~\cite{ba2015hardware,ba2016hardware}.
The challenge is that the filler cells used by these defenses must be tamper-evident, i.e., a defender must be able to detect if an attacker removed filler cells to implement their Trojan.
Previous work shows that filling the entire component layer with tamper-evident filler cells (e.g.~\cite{xiao2013bisa}) is infeasible due to routing congestion~\cite{ba2015hardware}.
To make routing feasible, the most recent placement-centric defense focuses on filling the unused placement sites nearest security-critical logic first~\cite{ba2015hardware,ba2016hardware}.

Such placement-centric defenses increase the complexity faced by the attacker in two ways.
First, it is harder for the attacker to find \emph{contiguous unused placement sites} to implement their Trojan's logic.
Second, an indirect complication is increased \emph{intra-Trojan routing} complexity.
The more distributed the attacker's placement sites, the more long (i.e., uses upper routing layers) routes the attacker must create.
Additionally, since the unused placement sites are far away from security critical logic, the attacker must make a longer, more complex, route to connect their hardware Trojan to the victim security-critical wire.

%% file: sections/4_design.tex
Drawing from existing untrusted foundry defenses, we create a extensible set of IC layout attack metrics. We unify the objectives of existing defenses by decomposing the act of inserting a hardware Trojan into ICs at an untrusted foundry into three fundamental tasks and corresponding metrics:
\begin{enumerate}
  \item Trojan logic placement: \textbf{Trigger Space}
  \item Victim/Trojan integration: \textbf{Net Blockage}
  \item Intra-Trojan routing: \textbf{Route Distance}
\end{enumerate}  
These tasks and accompanying metrics are the foundation for our methodology of assessing defensive coverage of an IC layout against an untrusted foundry.
We implement our methodology as \icad{}.

\subsection{Challenges of Trojan Placement}
\label{section:trojan_placement_challenges}
The first phase of mounting a fabrication-time attack is Trojan placement. This requires locating unused placement sites on the placement grid to insert additional circuit components. While prior work~\cite{xiao2013bisa,ba2015hardware,ba2016hardware} employs the notion of limiting the quantity of unused placement sites as a defense against fabrication-time attacks, how can we characterize unused placement sites to gain insight into the feasibility of a fabrication-time attack on a given IC layout?

Only 60--70\% of the placement cites are occupied in a typical IC layout to allow space for routing~\cite{a2}. To facilitate Trojan routing, an attacker prefers open placement sites form contiguous (adjacent) regions. This allows the attacker to drop-in a pre-designed Trojan, or if one had not been pre-designed, it minimizes the intra-Trojan routing complexity by confining the intra-Trojan routing to the lowest routing layers, i.e., reducing the jumping and jogging of nets. Such adjacency is classified in image processing as ``4-connected''. Therefore, a key factor that determines the difficulty of mounting fabrication-time attacks is the difficulty of inserting additional circuit components into a finalized IC design. We rank this difficulty in increasing order as follows.

\begin{enumerate}
    \item \textbf{Trivial:} the Trojan components fit within a single contiguous group of 4-connected placement sites.
    \item \textbf{Difficult:} the Trojan components must be split across multiple contiguous groups of 4-connected placement sites. The more groups of placement site groups, the more difficult intra-Trojan routing becomes.
    \item \textbf{Not Possible:} the total area required by the hardware Trojan exceeds that of available placement sites.
\end{enumerate}

\begin{figure}[t]
\centering
\includegraphics[width=0.48\textwidth]{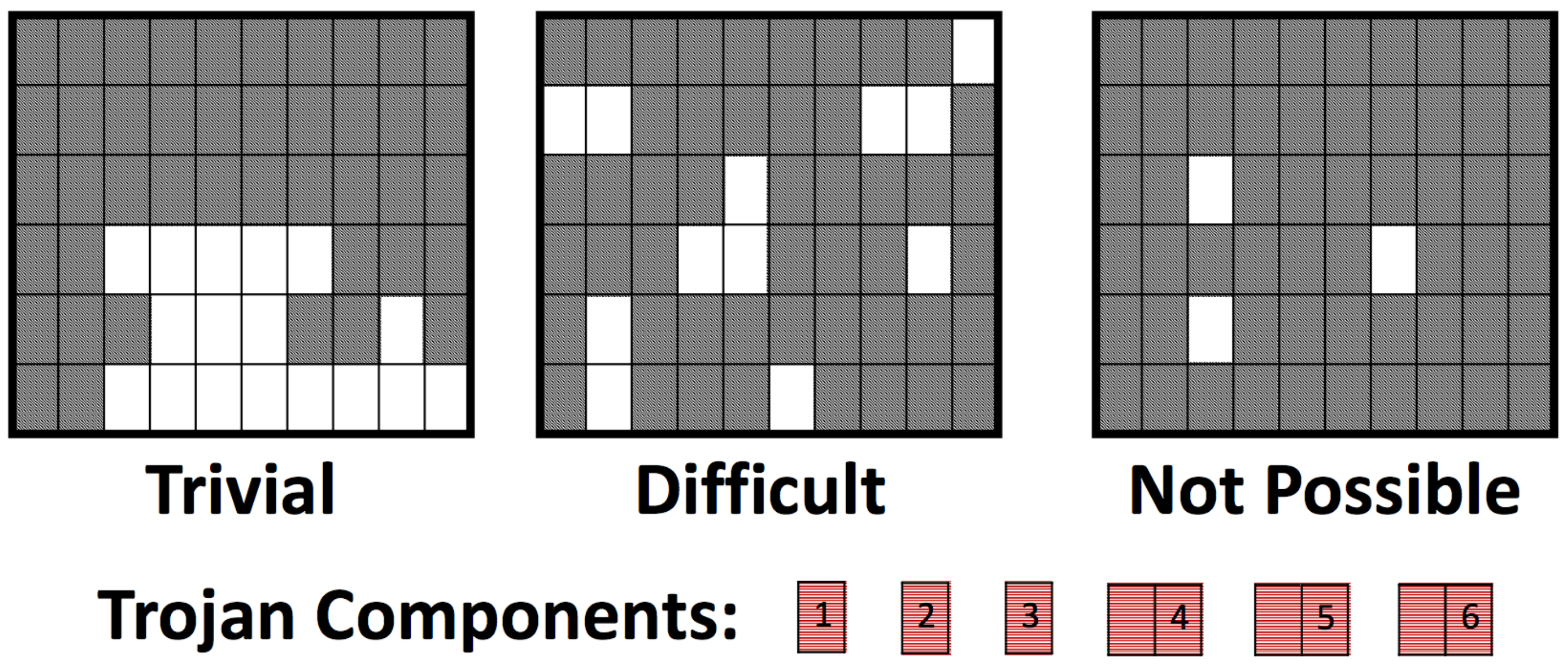}
\caption{\footnotesize Assume an attacker is attempting to insert 6 additional Trojan components that consume a total of 9 placement sites (as shown). If inserting these components on the \textit{Trivial} placement grid (left), they can be placed adjacent to each other to simplify intra-Trojan routing. If inserting these components on the \textit{Difficult} placement grid (middle), they must be scattered across the grid, making intra-Trojan routing more challenging. The \textit{Not Possible} placement grid (right) does not have enough empty placement sites to accommodate the Trojan components.}
\label{fig:open_space_difficulty}
\figline{}
\vspace*{-0.15in}
\end{figure}

Figure~\ref{fig:open_space_difficulty} illustrates these difficulty levels. The susceptibility of an IC design to fabrication-time attack can therefore be partially quantified by the size and number of contiguous open sites on the placement grid. This is the basis for \icadplural{} \textit{Trigger Space} metric.

\subsection{Challenges of Victim/Trojan Integration}
\label{section:victimtrojan}

Routing the Trojan payload to the targeted security-critical net requires the attacker to locate the nets of interest in the IC layout.
We assume the worst case: the attacker has knowledge of all security-critical nets in the design, particularly, the nets they are trying to extract information from or influence.
An example of such a net in the OR1200 processor~\cite{or1200} is the net that holds the privilege bit.
The attacker can acquire this knowledge either through a design-phase co-conspirator or through advanced reverse-engineering techniques~\cite{a2}.
No matter how the attacker gains this information, we assume they have it with zero additional effort.

We extend this threat to include nets that influence security-critical nets.
To increase stealth, an attacker could also trace backwards from the targeted security-critical net, through logic gates, to identify nets that influence the value of the targeted security-critical net.
This is called the \textit{fan-in} of the targeted net.
By connecting in this way, the attacker sacrifices controllability for stealth as their circuit modification is now physically separated from the security-critical net.
To gain back controllability, attackers must create a more complex (hence larger) trigger circuit---decreasing the Trigger Space score, as well as increasing the likelihood of visual and/or side-channel detection.
This tradeoff limits how many levels back the attacker can integrate their payload.

\begin{figure}[t]
\centering
\includegraphics[width=0.48\textwidth]{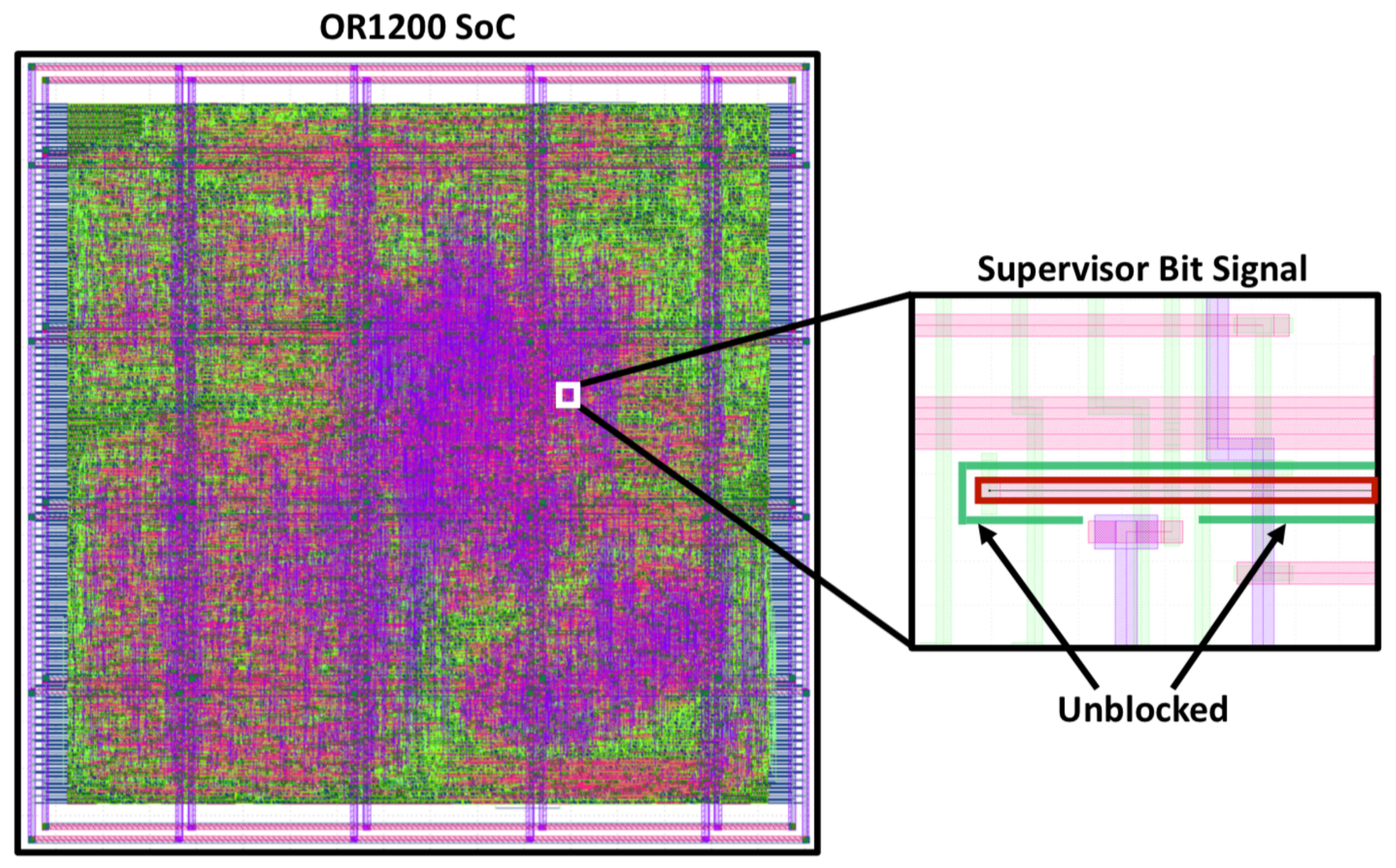}
\caption{\footnotesize The supervisor bit signal of the OR1200 processor SoC is the data input to the supervisor register of the OR1200 CPU. The supervisor register stores the privilege mode the processor is currently executing in. Changing the value on this net changes the privilege level of the processor allowing an attacker to execute privileged instructions. The more congested the area around this net, the more difficult it is for a foundry-level attacker to attach (or route in close proximity) a rogue wire to it.}
\label{fig:net_blockage}
\figline{}
\vspace*{-0.15in}
\end{figure}

No matter if the attacker is attacking the targeted security-critical wire directly or indirectly, the attacker must attach to some victim wire or route directly adjacent to it.
Since an IC layout is three-dimensional, it is possible for the attacker to attach to any open point on the victim wire, either on the same layer (i.e., North, South, East, West) or by coming in from an adjacent layer (i.e., above or below).
In the worst case, there are no other nets blocking the attacker from attaching to the targeted security-critical net or its $N$-level-deep influencers.
In the best case, all attachment points are blocked by other nets.
To quantify the number of points along, above, and below a targeted security-critical wire---and its $N$-deep fan-in---we implement the \textit{Net Blockage} metric.
Figure~\ref{fig:net_blockage} shows the open (unblocked) integration points for the privilege net on the OR1200 processor.

\subsection{Challenges of Intra-Trojan Routing}
\label{section:intratrojan}

The final phase of a fabrication-time attack is Intra-Trojan routing.
Intra-Trojan routing requires connecting the components that comprise the trigger and payload portions of the hardware Trojan together---including connecting to the integration point with the victim---to form a complete hardware Trojan.
In the worst case, the attacker is able to find a single contiguous region to place the trigger and payload components that is nearby the victim security-critical net.
Thus, routing the trigger and payload components will be trivial and the wire used to inject the payload will be short.
In the best case, the attacker will have to implement their attack using many 4-connected placement regions (i.e., low Trigger Space score) and the only integration point on the targeted security-critical net (i.e., high Net Blockage score) is as far away from the open placement regions.
Hence, we focus on quantifying the difficulty of routing the payload output to open attachment points on targeted security-critical nets (and its $N$-deep fan-in).
To this end, we identify two challenges of intra-Trojan routing:
\begin{itemize}
     \item Comply with design and fabrication rules
     \item Meet Trojan and payload-delivery timing requirements
\end{itemize}

\textbf{Complying with Design Rules.} For each process technology, there are many rules associated with how wires and components must be laid out in a design. Some of these rules are defined in the Library Exchange Format (LEF)~\cite{lef_def_format} and contained in files that are loaded by modern Computer Aided Design (CAD) tools throughout the IC design process. There are two types of design rules: 1) those regarding the construction of circuit components (i.e., standard cells), and 2) those regarding routing. We classify these as \textit{component design rules} and \textit{routing design rules}, respectively. As technology nodes shrink, both sets rules are becoming increasingly complex~\cite{dr_complexity_rising}.

It is vital for an attacker to comply with these design rules as violating them risks exposure. If an attacker inserts additional logic gates (standard cells) by making copies of existing components in a design, they can avoid violating \textit{component design rules} involved with Trojan placement. However, to connect a wire from the Trojan payload to security-critical target net(s), they must perform custom Trojan routing. Therefore, complying with \textit{routing design rules} is a concern. Routing design rules include specifications for the minimum distance between two nets on a specific routing layer, the minimum width of nets on a given layer, etc. Complying with these rules becomes easier for an attacker if the security-critical target net(s) are not blocked by other wires or components. The higher the Net Blockage score, the more difficult it is to make a connection, the more complex---and error prone---the route.

\textbf{Meeting Timing Requirements.} Every wire in an IC has a resistance and a capacitance, making it behave like an RC circuit, i.e., there is a time delay associated with driving the wire $high$ (logic $1$) or $low$ (logic $0$). The longer the wire, the more time delay there is~\cite{elmore1948transient}. If the target net(s) has timing constraints (e.g., setup and hold times) that dictate when the payload signal must arrive at the target net for the attack to be successful, the Trojan routing must meet these constraints. Furthermore, the farther the target net is from the payload circuit, the more obstacles that must be routed around, increasing the routing distance even further. This is the basis for \icadplural{} \textit{Route Distance} metric. A natural limit for Route Distance is dictated by the clock frequency of the victim circuit, as most attacks must operate synchronously with their victim.

%% file: sections/5_implementation.tex
\begin{figure}[t]
\centering
\includegraphics[width=0.9\columnwidth]{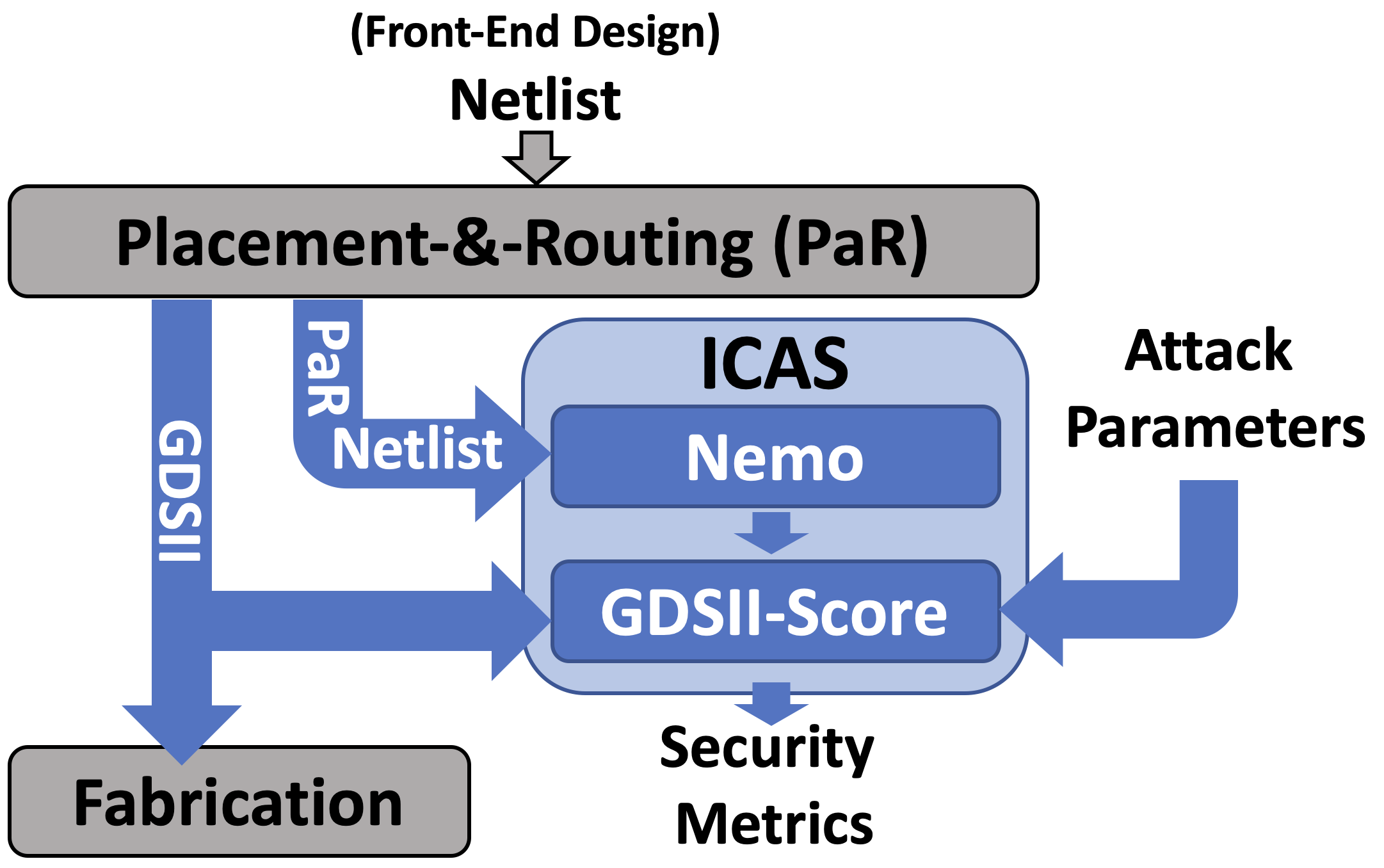}
\caption{\footnotesize \icad{} consists of two tools, {\bf Nemo} and {\bf GDSII-Score}, and fits into the existing IC design process (Fig.~\ref{fig:ic_design_flow}) between PaR and fabrication. Nemo analyzes a gate-level netlist and traces the fan-in to security-critical nets in a design. GDSII-Score analyzes a GDSII file (i.e., an IC layout) and computes metrics quantifying its vulnerability to a set of foundry-level attacks.}
\label{fig:icad_architecture}
\figline{}
\vspace*{-0.15in}
\end{figure}

The \icad{} framework is comprised of two tools, {\bf Nemo} and {\bf GDSII-Score}, as shown in Figure~\ref{fig:icad_architecture}. Nemo identifies security-critical wires based on designer annotations and circuit dataflow, while GDSII-Score assess the defensive coverage of a given IC layout against a set of attacks. \icad{} takes as input four sets of files: 1) gate-level netlist (generated \emph{after} all physical layout optimizations), 2) process technology files, 3) physical layout files, and 4) set of attacks. The process technology files include a Library Exchange Format (LEF) file and layer map file~\cite{lef_def_format,layer_map_format}. The physical layout files include a Design Exchange Format (DEF) file and the GDSII file of an IC layout~\cite{gdsii_format,lef_def_format}. The attack files are are a list of properties for each attack to assess coverage against: number of transistors, security-critical wire(s) to attach to, and timing constraints. All \icad{} input files except the attack files are either generated-by or inputs-to the back-end IC design phase, and hence are readily available to back-end designers.

Though \icad{} is extensible, our implementation includes three security metrics that capture the challenges faced by a foundry-level attacker looking to insert a hardware Trojan: amount and size of open-placement regions (Trigger Space), quantity of viable attachment points to targeted security-critical (and influencer) nets (Net Blockage), and the proximity of open placement regions to targeted security-critical net(s) (Route Distance). Together with the attack requirements, these metrics quantify the complexity an attacker faces for each step of inserting specific hardware Trojans into the given IC layout. We describe the implementation of both \icad{} components below.

\subsection{Nemo}\label{section:implementation_nemo}
Nemo is the first analysis tool in the \icad{} framework. It takes as input a Verilog netlist and automatically identifies the fan-in to root security-critical nets, which is output in the form of a Graphviz dot file~\cite{ellson2004graphviz}. This is necessary since the inter-connected nature of signals within a circuit design means an adversary could influence the state of security-critical nets by controlling a net that is a part of its fan-in. Like prior work~\cite{jin2010dftt,linscott2018swan}, Nemo assumes that HDL designers have appended a unique signal prefix to various signals considered ``security-critical'' in their designs at the RTL level. We make this assumption since determining what signals are ``security critical'' is easier with the semantics provided at the behavioral (RTL) level. For annotation, we leverage existing security-critical signal identification techniques~\cite{jin2010dftt,ba2015hardware,specs15,zhang2017identifying}. Unfortunately, existing tools do not extend past the RTL design phase. Thus, Nemo's task is to bridge the semantic gap and uncover duplicated or renamed security-critical signals in the post-PaR netlist. Fortunately, while synthesis and layout tools do modify a netlist by duplicating and removing signals and components (as part of optimization and meeting performance requirements), they do not completely rename existing signals. This makes it possible for Nemo to identify root security-critical signals (flagged at the behavioral level) by name at the physical level. To avoid removal of security-critical signals, we modify synthesis and layout scripts to essentially lock them in place. Nemo works backwards from root security-critical signals to identify the fan-in to these signals. The search depth is a configurable parameter of Nemo.

Nemo is implemented as a back-end target module to the open-source Icarus Verilog (IVL)~\cite{icarus} Verilog compiler and simulation tool written in C++. The IVL front-end exposes an API to allow third-parties to develop custom back-end target modules. Nemo is a custom target module (also written in C++) designed to be loaded by IVL. Since gate-level netlists are often described with the same HDL that was synthesized to generate the netlist (e.g., Verilog), we utilize the IVL front-end to interpret the Verilog representation of the netlist and our custom back-end target module, Nemo, to analyze the netlist. We open-source Nemo~\cite{anon} and release instructions on how to compile and integrate Nemo with IVL.

\subsection{GDSII-Score}
\label{section:implementation_gds2score}

GDSII-Score is the second analysis tool in the \icad{} framework. GDSII-Score is an extensible Python framework for computing security metrics of a physical IC layout. It takes as input the following: Nemo output, GDSII file, DEF file, technology files (LEF and layer-map files), and attacks description file. First, GDSII-Score loads all input files and locates the security-critical nets within the physical layout. Next, it computes security metrics characterizing the susceptibility of an IC design to each of the input attacks. Specifically, the three security metrics that we implement are: \textbf{Trigger Space}: the difficulty of implementing the hardware Trojan, \textbf{Net Blockage}: the difficulty of Trojan/victim integration, and \textbf{Route Distance}: the difficulty of meeting Trojan timing constraints.
We open source the GDSII-Score framework and our security metric implementations~\cite{anon}.

\subsubsection{\textbf{Metric 1: Trigger Space.}}The Trigger Space metric quantifies the challenges of Trojan placement (\S~\ref{section:trojan_placement_challenges}).
It computes a histogram of open 4-connected regions of all sizes on an IC's placement grid.
The more large 4-connected open placement regions available, the easier it is for an attacker to locate a space to insert additional Trojan circuit components at fabrication time.
A placement site is considered to be ``open'' if the site is empty, or if it is occupied by a filler cell.
Filler cells, or capacitor cells, are inserted into empty spaces during the last phase of layout to aid fabrication.
Since they are inactive, an attacker can create empty placement sites by removing them, without altering the functionality or timing characteristics of the victim IC.

To compute the trigger space histogram, GDSII-Score first constructs a bitmap representing the placement grid.
Placement sites occupied by standard cells (e.g., NAND gate transistors) are colored while those that are open are not.
Information about the size of the placement grid and the occupancy of each site in the grid is available in the Design Exchange Format (DEF) file produced by commercial PaR tools.
GDSII-Score then employs a breadth-first search algorithm to enumerate the maximum size of all 4-connected open placement regions.

\subsubsection{\textbf{Metric 2: Net Blockage.}} The Net Blockage metric quantifies the challenges of integrating the hardware Trojan's payload into the victim circuit (\S~\ref{section:victimtrojan}).
It computes the percent blockage around security-critical nets and their influencers.
The more congested the area surrounding security-critical nets, the more difficult it is to attach the Trojan circuitry to these nets.
There are two types of net blockage that are calculated for each security-critical net: \textit{same-layer} and \textit{adjacent-layer}.

\begin{figure}[t]
\centering
\includegraphics[width=0.70\columnwidth]{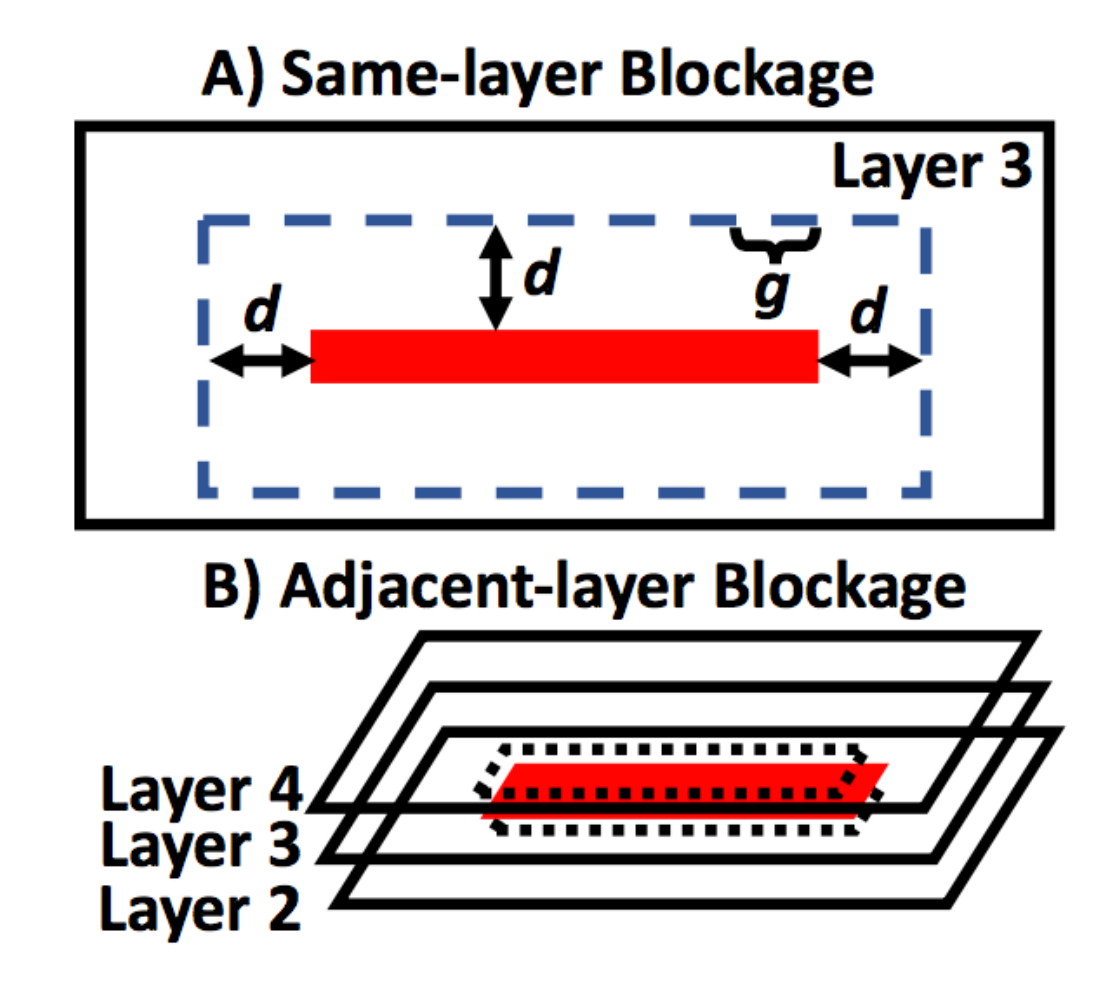}
\caption{\footnotesize A) Same-layer net blockage is computed by traversing the perimeter of the security-critical net, with granularity $g$, and extension distance $d$, and determining if such points lie inside another component in the layout. B) Adjacent-layer net blockage is computed by projecting the area of the security-critical net to the layers above and below and determining the area of the projections that are occupied by other components.}
\label{fig:net_blockage_algo}
\figline{}
\vspace*{-0.15in}
\end{figure}

Same-layer blockage is computed by traversing points around the perimeter (North, South, East, West) at a granularity of $g$, at a specific distance, $d$, around the security-critical net and determining which points lie within other circuit components, as detailed in Figure~\ref{fig:net_blockage_algo}a.
To determine if a specific point along the perimeter lies within the bounds of another circuit component, we utilize the point-in-polygon ray-casting algorithm~\cite{hughes2014computer}.
The extension distance, $d$, around the security-critical path element and the granularity of the perimeter traversal, $g$, are configurable in our implementation.
However, we default to an extension distance of one wire-pitch and a granularity of 1 database units, respectively, as defined in the process technology's LEF file.
The IC designs used in our evaluation are built using a 45\,$nm$ process technology, for which 1 database units is equivalent to 0.5\,$nm$.
Additionally, an open region is considered ``blocked'' if it is not wide enough for a minimal width wire to be routed through while maintaining the minimal amount of wire spacing required on that metal layer, as defined in the LEF file.
The percentage of the perimeter length that is blocked by other circuit components is considered the same-layer blockage percentage. 

Adjacent-layer blockage is computed by analyzing the area directly above and below a security-critical net, and computing the total area of overlap between other components, as detailed in Figure~\ref{fig:net_blockage_algo}b.
To calculate this overlap area we utilize the Weiler-Atherton polygon clipping algorithm~\cite{weiler_atherton}.
Additionally, any un-blocked regions above or below the security-critical net are considered ``blocked'' if they are not large enough to accommodate the smallest possible via geometry allowed on the respective via layer, as defined in the LEF file.
The percentage of the total top and bottom area that is blocked by nearby circuit components is the adjacent-layer blockage percentage.

The same-layer and adjacent-layer blockage percentages are combined via a weighted average to form a comprehensive \textit{overall} net blockage percentage where $66\%$ is based on same-layer blockage (north, south, east, and west) and $33\%$ is based on adjacent-layer blockage (top and bottom). We weight the same-layer blockage by $66\%$, or $\frac{2}{3}$, because 4 out of 6 total sides of a wire (\textbf{north}, \textbf{south}, \textbf{east}, \textbf{west}, top, and bottom) are on the same layer. Likewise, we weight the adjacent-layer blockage by $33\%$, or $\frac{1}{3}$. 

Lastly, a total \textit{same-layer}, \textit{adjacent-layer}, and \textit{overall} net blockage metric is computed for the entire IC design.
For an IC design with $n$ security-critical nets, the \textit{same-layer} ($b_{\mbox{\footnotesize same}}$), \textit{adjacent-layer} ($b_{\mbox{\footnotesize adjacent}}$), and \textit{overall} ($b_{\mbox{\footnotesize overall}}$) net blockage metrics are computed according to equations~\ref{eq:total_sl_net_blockage}, \ref{eq:total_al_net_blockage}, and \ref{eq:total_overall_net_blockage}, respectively.

\begin{equation}
\label{eq:total_sl_net_blockage}
    b_{\mbox{\footnotesize same}} = \frac{\sum_{i=1}^{n} perimeter\_blocked_n}{\sum_{i=1}^{n} perimeter_n}
\end{equation}

\begin{equation}
\label{eq:total_al_net_blockage}
    b_{\mbox{\footnotesize adjacent}} = \frac{\sum_{i=1}^{n} area\_blocked_n}{\sum_{i=1}^{n} 2 * area_n}
\end{equation}

\begin{equation}
\label{eq:total_overall_net_blockage}
    b_{\mbox{\footnotesize overall}} = \left( \frac{2}{3} * b_{\mbox{\footnotesize same}} \right) + \left( \frac{1}{3} * b_{\mbox{\footnotesize adjacent}} \right)
\end{equation}

\subsubsection{\textbf{Metric 3: Route Distance.}}
The Route Distance metric combines the Net blockage and Trigger Space metrics to quantify the difficulty of of meeting Trojan and attack timing constraints (\S~\ref{section:intratrojan}).
It computes a conservative estimate, i.e., Manhattan distance, for the minimal routing distance between open trigger placement sites and the $n$ least blocked integration sites on the targeted security critical nets.
It cross-references each Manhattan distance with the distribution of net lengths within the entire IC design.
Net length can impact whether or not the Trojan circuit will meet timing constraints and function properly.
Understanding where in the distribution of net lengths the Trojan routing falls provides insights into the ability of the Trojan circuit to meet its timing requirements and is an opportunity for outlier-based defenses.
In summary, the more Manhattan distances that fall within one standard deviation of the mean net length, the easier it is to carry out an attack.

We implement the Route Distance metric as follows.
First, the Net Blockage and Trigger Space metrics are computed.
Next, the the distribution of all net-lengths within the IC layout are computed.
Then, two-dimensional Manhattan distances between all unblocked nets ($<100\%$ overall net blockage) and trigger spaces are calculated.
The Manhattan distance calculated is the minimum distance between a given trigger space and security-critical net, i.e., the minimum distance between any placement site within the given trigger space and any unblocked location on the targeted security-critical net.
Lastly, each Manhattan distance is reported in terms of standard deviations away from the mean net-length in the given IC layout.

%% file: sections/6_evaluation.tex
We use \icad{} to quantify the defensive coverage of existing defensive layout techniques---revealing that gaps persist.
First, we analyze the effectiveness of undirected defenses~\cite{xiao2013bisa}.
Specifically, we measure the impact of varying both physical and electrical back-end design parameters of the same IC layout on its susceptibility to attack.
Second, we analyze the effectiveness of directed defenses~\cite{ba2015hardware,ba2016hardware}.
Specifically, we measure the coverage of existing, placement-oriented, defensive layout schemes in preventing the insertion of an attack by the foundry.
Beyond revealing gaps, our results reveal that there is an opportunity for a improving both directed and undirected defenses that systematically eliminates Trojan/victim integration points.
Lastly, our evaluation also demonstrates that \icad{} is design-agnostic, works with commercial tools, and scales to complex IC layouts.

\input{sections/trojan_designs_table}

\begin{figure*}[!t]
\centering
\includegraphics[width=0.78\textwidth]{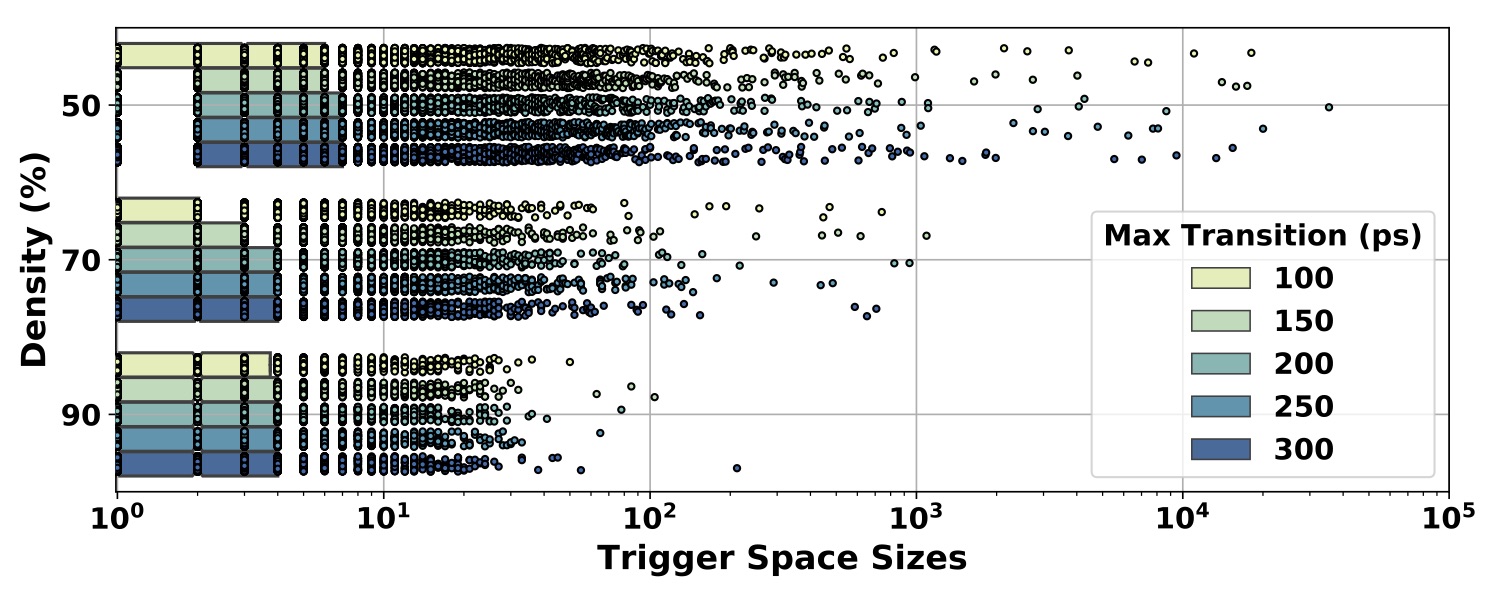}
\vspace*{-0.15in}
\caption{\footnotesize Trigger Space distributions for 15 different OR1200 processor IC layouts. Core density and max transition time parameters are varied across the layouts, while target clock frequency is held constant at 1 $GHz$. The boxes represent the middle 50\% (interquartile range or IQR) of open placement regions in a given layout, while the dots represent individual open placement region sizes.}
\label{fig:eval_2_ts_plot}
\figline{}
\vspace*{-0.15in}
\end{figure*}

\subsection{Experimental Setup} 
\label{section:eval_exp_setup}
We utilize three IC designs for our evaluations: \textit{OR1200 processor SoC}, \textit{AES accelerator}, and \textit{DSP accelerator}. The OR1200 processor SoC is an open-source design~\cite{or1200} used in previous fabrication-time attack studies~\cite{a2}. The AES and DSP accelerator designs are open-sourced under the Common Evaluation Platform (CEP) benchmark suite~\cite{cep}. The OR1200 processor SoC consists of a 5-stage pipelined OR1200 CPU that implements the 32-bit OR1K instruction set and Wishbone bus interface. The AES accelerator supports 128-bit key sizes. The DSP accelerator implements a Fast Fourier Transform (FFT) algorithm. 

All designs target a 45$nm$ Silicon-On-Insulator (SOI) process technology.
We synthesize and place-and-route all designs with Cadence Genus (version 16.23) and Innovus (version 17.1), respectively.
In our first evaluation (\S~\ref{section:eval_1}) the design constraints (clock frequency, max transition time, core density) used for both synthesis and layout are varied as noted.
However, in our second evaluation (\S~\ref{section:eval_2}) the same design constraints (100 $MHz$ clock frequency, 100 $ps$ max transition time, 60\% core density) were used for both synthesis and layout to form a common baseline.
All ICs are synthesized and placed-and-routed on a server with 2.5 $GHz$ Intel Xeon E5-2640 CPU and 64 $GB$ of memory running Red Hat Enterprise Linux (version 6.9). 

\subsubsection{\textbf{Security-critical Signals}}

The first tool in the \icad{} flow is Nemo.
Nemo tracks security-critical signals from the HDL level to the IC layout level.
The first step is flagging root security-critical signals at the RTL level, for each IC design.
For the OR1200 processor SoC, the supervisor bit signal $supv$ is flagged.
We select this signal because one can alter the state of this bit to escalate the privilege mode of the processor~\cite{a2}.
For the AES accelerator, we flag all 128 key bits as security-critical.
The $next\_out$ signal within the DSP accelerator was flagged as security-critical.
The $next\_out$ signal of the DSP accelerator indicates to external hardware when an FFT computation is ready at the output registers.
Tampering with the $next\_out$ signal allows the attacker to hide specific outputs of the DSP accelerator.
Lastly, Nemo marks, for each design's IC layout, all root security-critical nets and their 2-deep fan-in as security-critical nets.

\subsubsection{\textbf{Hardware Trojans}}

 Table~\ref{table:trojan_designs} lists the hardware Trojan designs that we use in our quantification of defensive coverage.
 The first two Trojan designs (analog and digital variants of A2) are attacks on the OR1200 processor and DSP accelerator ICs.
 With respect to the OR1200, the A2 attacks act as a hardware foothold~\cite{king08} for a software-level privilege escalation attack.
 With respect to the DSP accelerator, the A2 attacks suppress the \textit{next\_out} signal (\S~\ref{section:eval_exp_setup}).
 The Privilege Escalation Trojan targets solely the OR1200 and the Key Leak solely the AES accelerator.

\subsubsection{\textbf{Build Environment}}

Both \icad{} tools (Nemo and GDSII-Score) were run on the same server as the synthesis and place-and-route CAD tools.
Nemo and Icarus Verilog were compiled from source using GCC (version 4.4.7).
For increased performance, GDSII-Score was executed using the PyPy Python interpreter with JIT compiler (version 4.0.1)~\cite{pypy}.

\subsection{Undirected Defense Coverage}
\label{section:eval_1}

As detailed in \S~\ref{sec:undirected}, a defensive strategy for protecting an IC layout from foundry-level attackers is to exploit physical layout parameters (e.g., core density, clock frequency, and max transition time) offered by commercial CAD tools to increase congestion---hopefully around security-critical wires.
The tradeoff is that while this is a low cost defense in that CAD tools already expose such knobs, the entire design is impacted and there is no guarantee that security-critical wires will be protected.
We use \icad{} and its three security metrics to quantify the effectiveness of such undirected approaches~\cite{xiao2013bisa}.

To uncover the impact of each parameter, we start by generating 60 different physical layouts of the OR1200 processor design by varying:
\begin{enumerate}
 \item \textbf{Target Core Density} (\%): 50, 70, 90
 \item \textbf{Clock Frequency} ($MHz$): 100, 200, 500, 1000
 \item \textbf{Max Transition Time} ($ps$): 100, 150, 200, 250, 300
\end{enumerate}
Target core density is a measure of how congested the placement grid is.
Typically, designers select die dimensions that achieve $\sim$60--70\% placement density to allow space for routing~\cite{a2}.
Target clock frequency is the desired speed at which the circuitry should perform.
Typically, designers select the clock frequency based on performance goals.
Max transition time is the longest time required for the driving pin of a net to change logical values.
Typically, designers choose a value for max transition time based upon power consumption and combinational logic delay constraints.

For each of the 60 layout variations we compute \icad{} metrics.
Figures~\ref{fig:eval_2_ts_plot}, \ref{fig:eval_2_nb_plot}, and~\ref{fig:eval_2_ed_plot} provide a visual representation for each metric.
Overlaid on Figure~\ref{fig:eval_2_ed_plot} are the number of unique attack (color-coded) implementations for each Trojan (Tab.~\ref{table:trojan_designs}) at six parameter configurations.
Across the 60 IC layouts, the time it took \icad{} to complete its analyses ranged from 38 seconds to 18 minutes.
On average, this translates to less than 10\% of the combined synthesize and place-and-route run-times.
These run-time results demonstrate the deployability of \icad{} as a back-end design analysis tool.
Overall, our evaluation indicates that while some of these layout parameters do increase attacker complexity, none are sufficient on their own.
Next we break down the results metric-by-metric.

\subsubsection{\textbf{Trigger Space Analysis.}} Figure~\ref{fig:eval_2_ts_plot} shows the distributions of open trigger spaces across 15 unique OR1200 layouts.
We vary target core density and max transition time parameters across layouts, while we fix the target clock frequency at 1\,GHz.
A trigger space is defined as a contiguous region of open placement sites on the device layer placement grid and is measured by number of contiguous ``4-connected'' placement sites.
Each box represents the middle 50\%, or interquartile range (IQR), of open trigger space sizes for a given IC layout.
The dots represent individual data points within and outside the IQR. Our empirical results affirm prior notions~\cite{xiao2013bisa,ba2015hardware,ba2016hardware} that increasing the target core density of an IC layout results in fewer large open spaces to insert hardware Trojans.
Additionally our results indicate that at lower densities, decreasing the max transition time constraint decreases the median trigger space size.
Similar trends occur at lower clock frequencies.
While results show that modulating target core density is effective, observe that even in the best case, large trigger spaces remain.

\subsubsection{\textbf{Net Blockage Analysis.}} 
\begin{figure}[!t]
\centering
\includegraphics[width=0.45\textwidth]{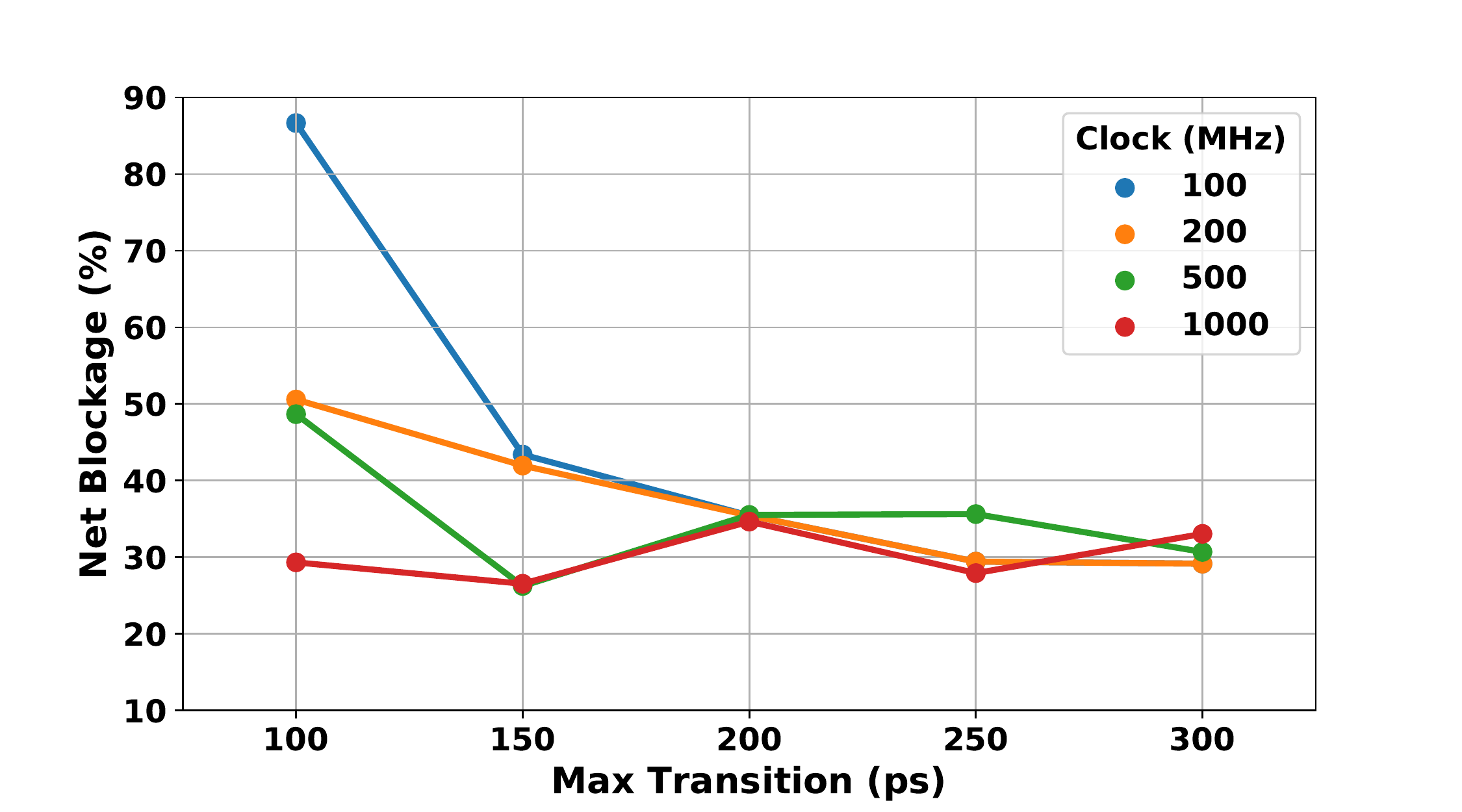}
\vspace*{-0.25in}
\caption{\footnotesize \textit{Overall} Net Blockage results computed across 20 different OR1200 processor IC layouts. A target density of 50\% was used for all layouts, while target clock frequency and max transition time parameters were varied.}
\label{fig:eval_2_nb_plot}
\figline{}
\vspace*{-0.15in}
\end{figure}

Figure~\ref{fig:eval_2_nb_plot} shows the Net Blockage metric (Eq.~\ref{eq:total_overall_net_blockage}) computed across 20 unique OR1200 layouts.
We fix the target density at 50\% across all layouts, while the target clock frequency and max transition time are varied (as listed above).
The results show that at lower clock frequencies a smaller max transition time parameter corresponds to increased Net Blockage.
This corresponds to less open Trojan/victim integration points available to the attacker.
However, as clock speed increases, the correlation between max transition time and overall Net Blockage deteriorates.
Intuitively, smaller max transition times should lead to smaller average net-lengths within the design, as transition time is a function of the capacitive load on the net's driving pin~\cite{elmore1948transient}.
Shorter net-lengths result in more routing congestion as components cannot be spread-out across the die.
However, capacitive load (on a driving pin) is inversely proportional to frequency, thus at higher clock frequencies the max-transition time constraint is more easily satisfied, and altering it has less effect on the Net Blockage.
Given these results, the effectiveness of modulating transition time is context dependent and---even in the best case---open integration points remain.

\subsubsection{\textbf{Route Distance Analysis.}}
\begin{figure}[!t]
\centering
\includegraphics[width=0.45\textwidth]{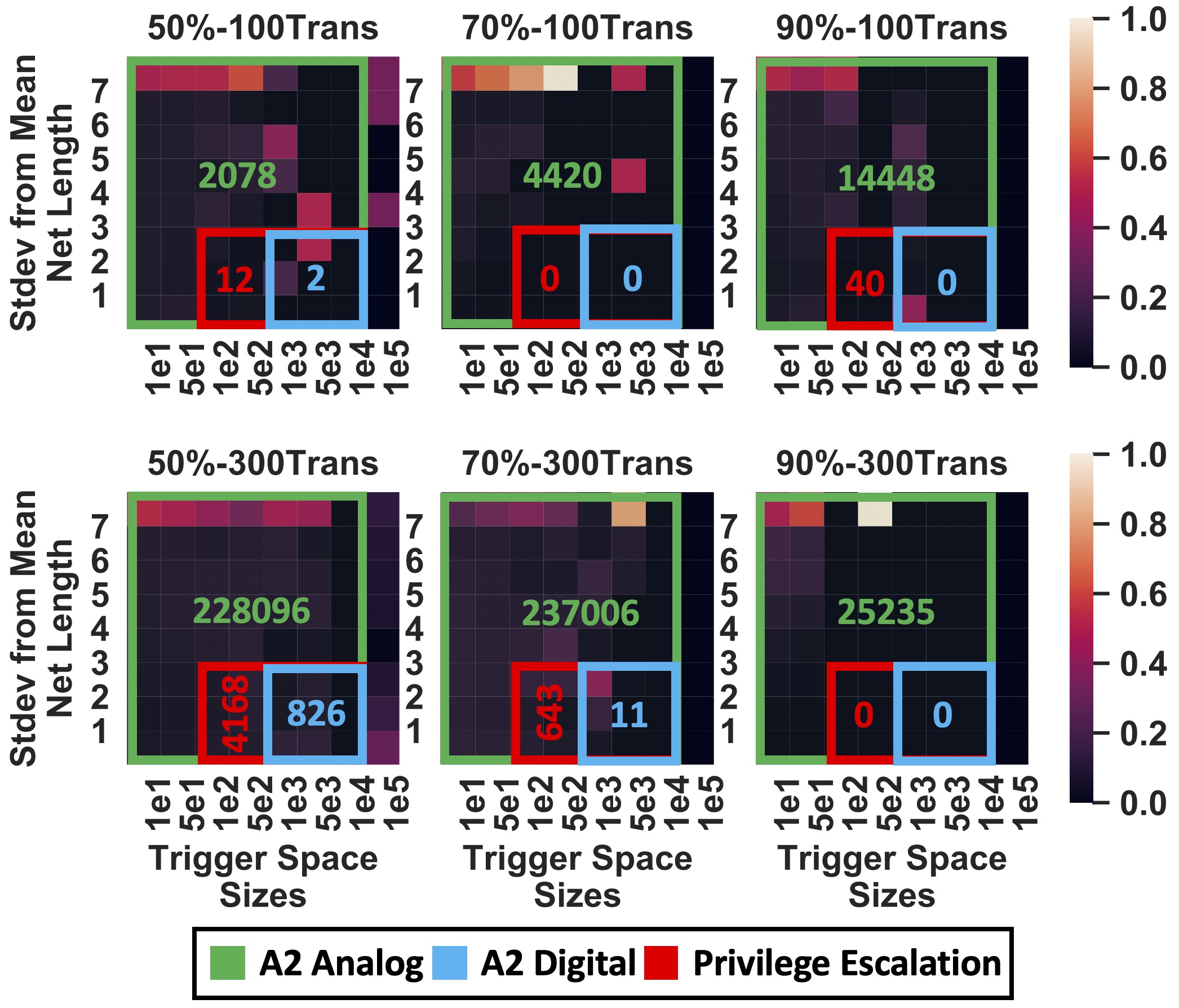}
\vspace*{-0.1in}
\caption{\footnotesize Heatmaps of routing distances across six unique IC layouts of the OR1200 processor. Core density and max transition times are labeled. Each heatmap is to be read column-wise, where each column is a histogram, i.e, the color intensity within a heatmap column indicates the percentage of (critical-net, trigger-space) pairs that are within a (y-axis) distance apart. 
Overlaid are rectangles, indicating regions on each heatmap a given attack can exploit, and numbers indicating the number of unique attack implementations.
}
\label{fig:eval_2_ed_plot}
\figline{}
\vspace*{-0.15in}
\end{figure}

Figure~\ref{fig:eval_2_ed_plot} shows the Route Distances across 6 various OR1200 layouts in the form of heatmaps that capture the trade space between layout parameters. Core density and max transition times were varied across the layouts (indicated in the labels), while clock frequency was held constant at 100\,$MHz$.
Each heatmap describes several (column-wise) histograms of Route Distances in terms of standard deviations from the mean net length observed in that particular IC layout (y-axis).
The Route Distances reported are those between any unblocked security-critical nets, and trigger spaces large enough to hold an attack of a given size range (x-axis).
That is, the color intensities within in a given heatmap column indicate the percentage of (security-critical-net, trigger-space) pairs in that column that are within a range of distance apart.
Additionally, overlaid on each heatmap are rectangles indicating the region of the heatmap where a given attack (Tab.~\ref{table:trojan_designs}) can be implemented, and the number of possible attack configurations, (security-critical-net, trigger-space) pairs, that can be exploited.

If timing is critical to the operation of an attacker's desired Trojan, (critical-net, trigger-space) pairs with routing distances significantly greater than the average net length in the IC layout are less likely to be viable candidates for constructing hardware Trojans. 
IC layouts with few desirable (critical-net, trigger-space) pairs are much more time-consuming to attack.
Namely, the IC layouts with heatmaps that indicate a higher percentages of far-apart (critical-net, trigger-space) pairs, where the trigger spaces are small, are most secure.
From Figure~\ref{fig:eval_2_ed_plot}, we conclude that at high density, max transition time has little affect on IC layout security; while at lower densities, lower max transition time designs are more secure.
Similar trends exist across other layout parameters, as shown in Figures~\ref{fig:all_rd_50density}--\ref{fig:all_rd_90density} in Appendix~\ref{section:appendix_1}.

\subsubsection{\textbf{Cost of Varying Layout Parameters.}}
The results indicate that increasing core density is effective, but incomplete, and increasing clock frequency and decreasing max transition time is marginally effective and incomplete.
While tuning these parameters is low cost to the designer, there is a cost to the design in terms of complexity and power requirements.
We elucidate by discussing how varying each design parameter (density, clock frequency, and max transition time) impacts non-security characteristics of a circuit design.

While increasing core density to 90\% makes placing-and-routing a Trojan more difficult, it also makes placing-and-routing the rest of the design more challenging.
Specifically, it can become nearly impossible to meet timing closure for the entire design if there is not enough space within the core area to re-size cells and/or add additional buffer cells.
Depending on performance and security requirements, a layout engineer may choose to relax timing constraints in order to achieve a higher core density. Alternatively, a layout engineer may attempt to surround security-critical nets with areas of high densities, while maintaining a lower overall core density, as previously suggested~\cite{ba2015hardware,ba2016hardware}.

Decreasing the maximum transition time and increasing the clock speed of an entire circuit design makes it more difficult to place-and-route a functional Trojan that meets timing constraints, but also directly impacts the performance characteristics of the circuit.
Additionally it is important to note that max transition time is related to the clock frequency, so varying one without the other changes performance tolerances.
While increasing the performance of the design might increase security, it comes at the cost of increasing power consumption.
Depending on the power-consumption requirements of the design, it may be possible for a designer to over-constrain these parameters for added security. 

\subsection{Directed Defense Coverage}
\label{section:eval_2}

As an alternative to probabilistically adding impediments to the attacker inserting a hardware Trojan, recent works proposes a directed approach.
As detailed in \S~\ref{sec:directed}, placement-centric directed defenses~\cite{ba2015hardware,ba2016hardware} attempt to prevent the attacker from implementing their Trojan by occupying all open placement sites with tamper-evident filler cells.
The limitation with such defenses is that it is infeasible to fill \emph{all} open placement sites with tamper-evident logic~\cite{ba2015hardware}.
Thus, the defenses focus their filling near security-critical logic, leaving gaps near the periphery of the IC layout.
Whether these open placement sites near the periphery are sufficient to implement an attack is an open question.

The goal of this evaluation is to determine not only if it is still possible for a foundry-level attacker to insert a hardware Trojan, given placement-centric defenses, but to quantify the number of viable implementations available to the attacker---to act as a surrogate for attacker complexity.
For the evaluation, We use our three IC designs (OR1200 processor SoC, AES accelerator, and DSP accelerator).
For each design, we create two IC layouts: (1) unprotected and (2) protected.
For the protected IC layout, we use the latest placement-centric defense~\cite{ba2016hardware}; using the identified security-critical wires (\S~\ref{section:eval_exp_setup}) to direct the defense.
We layout all IC designs using the these parameters: target clock frequency of 100\,$MHz$, max transition time of 100\,$ps$, and a target core density of 60\%.

We then use \icad{} to asses the defensive coverage of each of the six IC layouts.
This analysis has two goals: (1) determine whether the IC is vulnerable to attack and (2) understand the impact of applying the defense.
We answer both questions in an attack-centric manner using the hardware Trojans in Table~\ref{table:trojan_designs} to asses defensive coverage against.
For each attack/IC layout combination we plot the number of (security-critical-net, trigger-space) pairs that could be used in implementing each Trojan.
A (security-critical-net, trigger-space) pair is considered a viable candidate for implementing a Trojan if:
\begin{enumerate}
    \item the trigger space size is at \textit{least as large} as the minimum number of placement sites required to implement the desired hardware Trojan design
    \item the security-critical net is less than 100\% blocked
    \item if the hardware Trojan is ``Timing-Critical'', i.e., it must function at the design's core operating frequency, then the distance between the trigger space and open integration point on the security-critical net must be $\leq 3$ standard deviations from average net length; otherwise, any distance is allowed.\footnote{Three standard deviations from the average net length is the threshold for Trojan to integration point routing with violating timing constraints, because it accounts for 99.7\% of the designs' wires---outliers tend to be power wires. For an exact calculation it is possible to extract parasitics for a target for a Trojan's route to determine if it violates timing constraints.}
\end{enumerate}

\begin{figure}[t]
\centering
\includegraphics[width=0.45\textwidth]{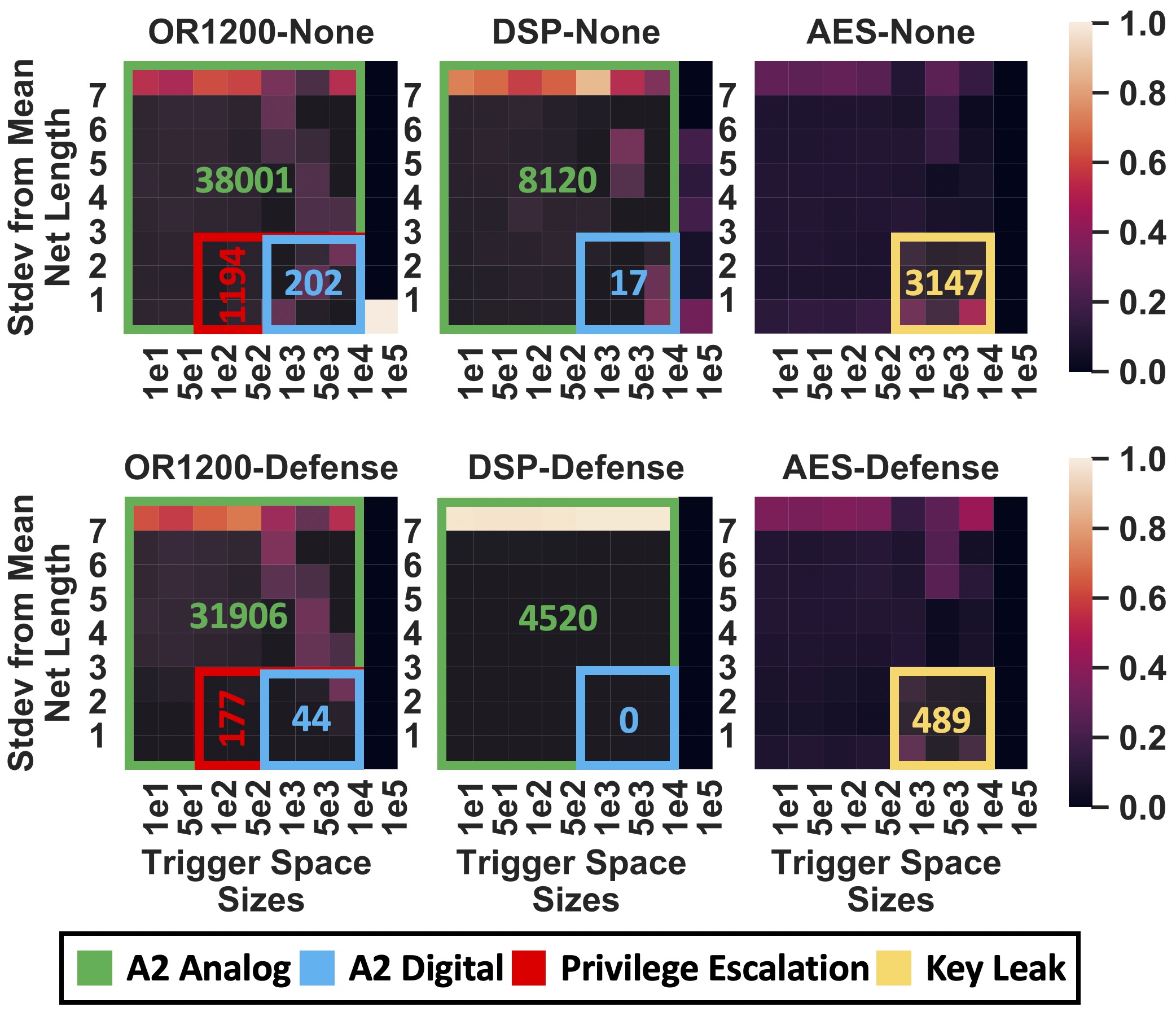}
\caption{\footnotesize Routing Distance heatmaps across three IC designs, with and without the placement-centric defense described in~\cite{ba2015hardware,ba2016hardware}. 
Heatmaps should be interpreted similar to Fig.~\ref{fig:eval_2_ed_plot}
}
\label{fig:eval_3_threat_assessment}
\figline{}
\vspace*{-0.15in}
\end{figure}

Figure~\ref{fig:eval_3_threat_assessment} shows the defensive coverage for each IC design.
Overlaid on each heatmap are rectangles (and numbers) indicating unique possible attack implementations.
These results show that existing placement-centric defenses are effective at \textit{reducing} an IC's fabrication-time attack surface, compared to no defense---but \textit{gaps persist}.
Given that filling placement sites with tamper-evident logic is already maximized, these results point to systematically adding congestion around security-critical wires as a means to close all remaining defensive gaps; i.e., a directed version with similar effect to existing undirected defenses.

%% file: sections/trojan_designs_table.tex
\begin{table}
\centering
\begin{tabular}{l c c c}
\multicolumn{1}{c}{\textbf{Trojan}} & 
\multicolumn{1}{c}{\textbf{\begin{tabular}[c]{@{}c@{}}\# Std\\ Cells\end{tabular}}}& 
\multicolumn{1}{c}{\textbf{\begin{tabular}[c]{@{}c@{}}\# Placement\\Sites\end{tabular}}}&
\multicolumn{1}{c}{\textbf{\begin{tabular}[c]{@{}c@{}}Timing\\Critical?\end{tabular}}}
\\ \hline \hline
\rowcolor{GreenHLight!30} A2 Analog~\cite{a2}                        &   2 &   20 & \xmark \\
\rowcolor{BlueHLight!30}A2 Digital~\cite{a2}                       &  91 & 1444 & \cmark \\
\rowcolor{RedHLight!30}\makecell[l]{Privilege\\Escalation~\cite{king08,hicks10}} &  25 &  342 & \cmark \\
\rowcolor{YellowHLight!30}Key Leak~\cite{trusthub}                   & 187 & 2553 & \cmark
\end{tabular}
\caption{\footnotesize Hardware Trojans used in defensive coverage assessment.}
\label{table:trojan_designs}
\vspace*{-0.32in}
\end{table}

%% file: sections/7_discussion.tex

\icad{} is the first tool to provide insights into the security of physical IC layouts. \icad{} is extensible across many dimensions including CAD tools, process technologies, security metrics, and fabrication-time attacks and defenses. To demonstrate \icadplural{} capabilities we implemented three security metrics (net blockage, trigger space, and routing distance) using it. The focus of this paper is using these metrics to quantitify the coverage of existing untrusted foundry defenses; which shows that IC designs are still vulnerable to attack. We envision uses for \icad{} beyond this, as an integral part of the IC design using commercial tools.

\textbf{\icad{}-Driven Defensive Layout:}
\icad{} provides an added notion of security to the IC layout process (place-and-route) to enable researchers to explore countermeasures against fabrication-time attacks. To the best of our knowledge, the targeted defensive IC layout techniques that exist~\cite{xiao2013bisa,ba2015hardware,ba2016hardware} are \textit{placement-centric}, i.e., filling unused space on the device layer with functional logic cells. While \icad{} is capable of evaluating placement-centric defensive layout techniques, its security-insights also asses \textit{routing-centric} defensive layout techniques. For example, layout engineers can leverage \icad{} to create high degrees of routing congestivity in close proximity to security-critical nets. \icadplural{} security metrics enable IC layout designers to optimize the security of both the placement \textit{and} routing of their designs.

\textbf{Extensibility of CAD Tools:} 
Almost all steps of the IC design process utilize CAD tools. \icad{} integrates into a commercial IC design process after placement-and-routing (Figure~\ref{fig:ic_design_flow}). While \icad{} is validated with IC layouts generated by Cadence tools, integrating \icad{} with other vendors' CAD tools requires no additional effort due to the common process technology (LEF) and GDSII specifications used by \icad{}. 


\textbf{Extensibility of Process Technologies:} 
We test \icad{} using IC layouts built with a 45\,$nm$ SOI process technology; however, \icad{} is agnostic of process technology. The LEF and layer map files (\S~\ref{section:implementation}) are the only \icad{} input files that are process technology dependent. A LEF file describes the geometries and characteristics of each standard cell in the cell library, and the layer map file describes the layer name-to-number mappings, respectively, for a given process technology. \icad{} adapts to different process technologies provided that all input files adhere to their specifications~\cite{lef_def_format,layer_map_format}.

\textbf{Extensibility of Security Metrics:}
GDSII-Score is the \icad{} tool that computes security metrics from an IC layout. It loads several files describing the IC layout to instantiate a single Python class (called ``Layout'') that contains query-able data structures containing a polygon representation of all components in the layout. Additionally, GDSII-Score contains several subroutines that compute spatial relationships between polygon objects and points within the layout. From these data structures and the provided subroutines, it is trivial to integrate additional novel metrics into GDSII-Score. To facilitate additional metrics, we open source GDSII-Score~\cite{anon}, and our three example metrics that demonstrate how to query the main ``Layout'' data structure.

%% file: sections/8_related_work.tex
Fabrication-time attacks and defenses have been extensively researched. Attacks have ranged in both size and triggering-complexity \cite{lin2009trojan,becker2013stealthy,shiyanovskii2010process,kumar2014parametric,a2}. Defenses against these attacks include: side-channel analysis~\cite{agrawal2007trojan,jin2008hardware,balasch2015electromagnetic,narasimhan2011tesr}, imaging~\cite{zhou2015detecting,adato2016rapid}, on-chip sensors~\cite{li2008speed,forte2013temperature}, and preventive measures~\cite{xiao2013bisa,cocchi2014circuit,ba2015hardware,ba2016hardware}. The most pertinent attacks and defenses are highlighted below.

\textbf{Untrusted-foundry Attacks:}
The first foundry-level attack was conceived by Lin {\em et al.}~\cite{lin2009trojan}. This hardware Trojan was comprised of approximately 100 additional logic gates and designed to covertly leak the keys of an AES cryptographic accelerator using spread spectrum communication to modulate information over a power side channel. While the authors only demonstrated this attack on an FPGA, they are the first to mention the possibility of this type of Trojan circuit being implanted at an untrusted foundry.

The A2 attack~\cite{a2} is the most recent fabrication-time attack. A2's analog triggering mechanism is stealthy, controllable, \emph{and} small. It prevents the Trojan from being exposed during post-fabrication testing, or unintentionally through common usage. The attack requires only two additional standard cells and evades every known detection mechanism to date. \icad{} quantifies the defensive coverage to these and other fabrication-time attacks. 

\textbf{Untrusted-foundry Defenses:} Most untrusted foundry defenses rely on post-fabrication \textit{detection} schemes~\cite{agrawal2007trojan,jin2008hardware,balasch2015electromagnetic,narasimhan2011tesr,zhou2015detecting,adato2016rapid,li2008speed,forte2013temperature}. \icad{} aims to guide innovation in \textit{preventive} defenses against fabrication-time attacks, for which few mechanisms currently exist~\cite{cocchi2014circuit,xiao2013bisa,ba2015hardware,ba2016hardware}. We highlight some of these preventive measures and how \icad{} could measure their effectiveness.

While preventive security-by-design was first explored at the behavioral (RTL) level by of Jin {\em et al.}~\cite{jin2010dftt}, Xiao {\em et al.} were the first to demonstrate security-by-design at the layout-level with their BISA (Built-In Self-Authentication) scheme~\cite{xiao2013bisa}. The \textit{undirected} BISA approach attempts to eliminate \emph{all} unused space on the device layer placement grid, and create routing congestion, by filling the device layer with interconnected tamper-resistant fill cells. Alternatively, recognizing the impracticality of filling 100\% of the empty placement sites in complex circuit designs, Ba {\em et al.} take a \textit{directed} approach to filling empty placement cites~\cite{ba2015hardware,ba2016hardware}. Specifically, they only fill empty placement sites in close proximity to security-critical nets.

%% file: sections/9_conclusion.tex
\icad{} is an extensible framework that we use to expose and quantify gaps in existing defenses to the threat posed by an untrusted foundry.
\icad{} has two high-level components: {\em Nemo}, a tool that bridges the semantic gap across IC design processes by tracking security-critical signals across all stages of hardware development and {\em GDSII-Score}, a tool that quantifies the difficulty a foundry-level attacker faces in attacking security-critical logic.
Experiments with over 60 IC layouts across three open-source hardware cores and four foundry-level hardware Trojans reveal that all current defenses leave the IC design vulnerable to attack---and some are totally ineffective.
These results show the value of a tool like \icad{} that can help designers identify and address defensive gaps.

From a high level, \icad{} is momentus in that it makes security a first-class concern during IC layout (in addition to power, area, and performance): \icad{} allows IC designers to measure the security implications of tool settings and design decisions. \icad{} fits well with existing IC design tools and flows, allowing them to consider security. \icad{} is a critical measurement tool that enables the systematic development of future physical-level defenses against the threat of an untrusted foundry.

%% file: sections/10_acknowledgement.tex
We thank Ted Lyszczarz, Brian Tyrrell, and other members of the MIT Lincoln Laboratory community for their thoughtful feedback that enhanced the quality of our work.

DISTRIBUTION STATEMENT A. Approved for public release. Distribution is unlimited. This material is based upon work supported by the Under Secretary of Defense for Research and Engineering under Air Force Contract No. FA8702-15-D-0001. Any opinions, findings, conclusions or recommendations expressed in this material are those of the author(s) and do not necessarily reflect the views of the Under Secretary of Defense for Research and Engineering.

This material is based upon work supported by the National Science Foundation Graduate Research Fellowship Program under Grant No. DGE 1256260. Any opinions, findings, and conclusions or recommendations expressed in this material are those of the author(s) and do not necessarily reflect the views of the National Science Foundation.

%% file: sections/11_appendix_1.tex
\noindent\begin{minipage}{\textwidth}
\centering
\makebox[\textwidth][c]{\includegraphics[width=1.1\textwidth]{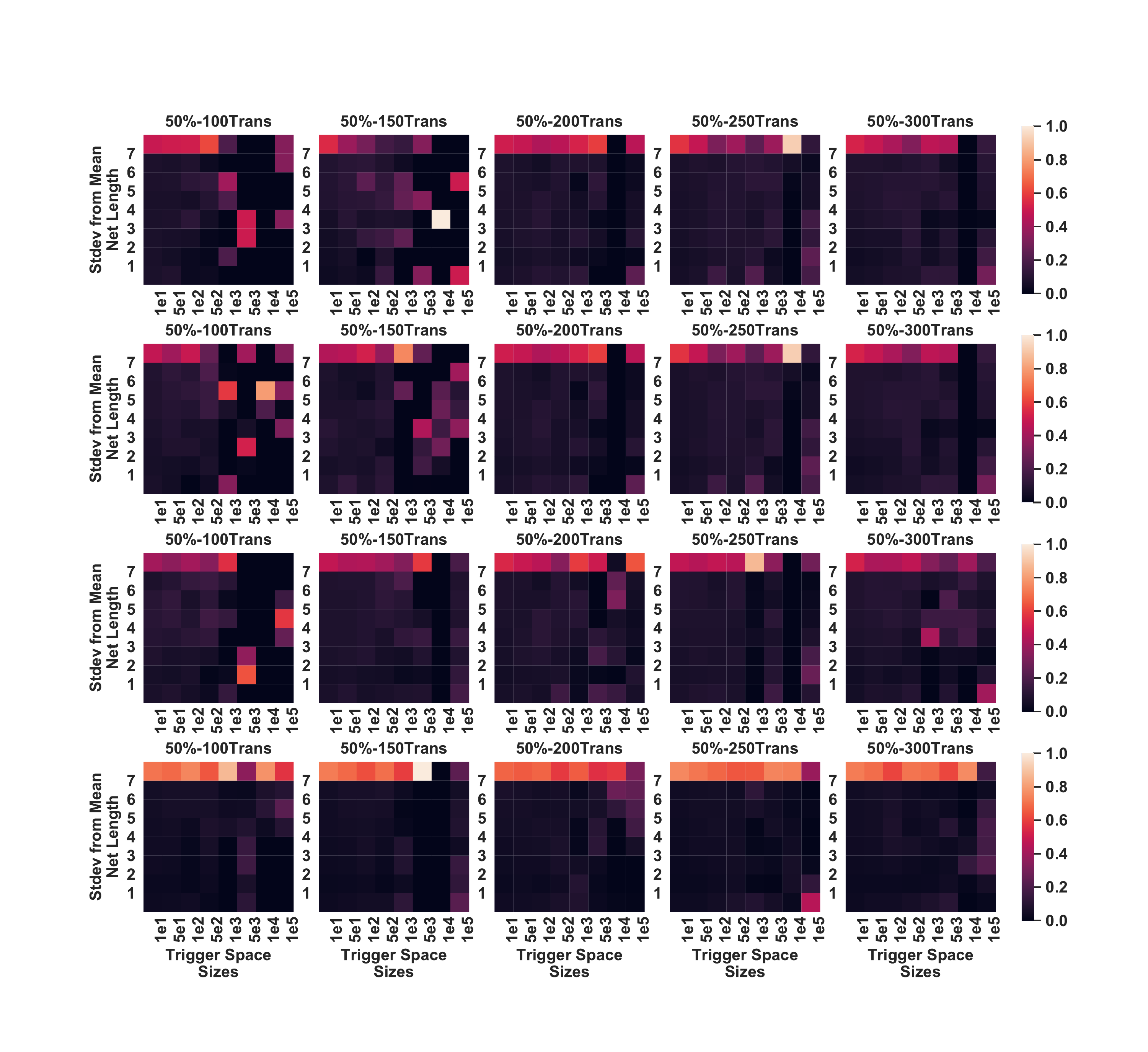}}
\vspace*{-0.08\textwidth}
\captionof{figure}{\footnotesize \textit{Route Distance Metric for OR1200 at 50\% Density).} A target density of 50\% was held across each layout, while target clock frequency and max transition time parameters were varied from 100\,MHz to 1000\,MHz and 100\,ps to 300\,ps respectively. Each heatmap is intended to be read column-wise, where each column is a histogram. The color intensity within a heatmap column indicates the percentage of (critical-net, trigger-space) pairs, within that column, that are within a range of distance away. The y-axis reports the distance in terms of standard deviations from the overall mean net-length in each design. The x-axis reports the trigger space sizes in number of contiguous placement sites. Designs with smaller trigger-spaces and long route distances are more resistant to fabrication-time attacks. Namely, a heatmap column that is completely dark indicates no (critical-net, trigger-space) pairs, or attack points, and a column that is completely dark except for the top-most cell is the second most secure.}
\label{fig:all_rd_50density}
\end{minipage}


\begin{figure*}
\centering
\makebox[\textwidth][c]{\includegraphics[width=1.1\textwidth]{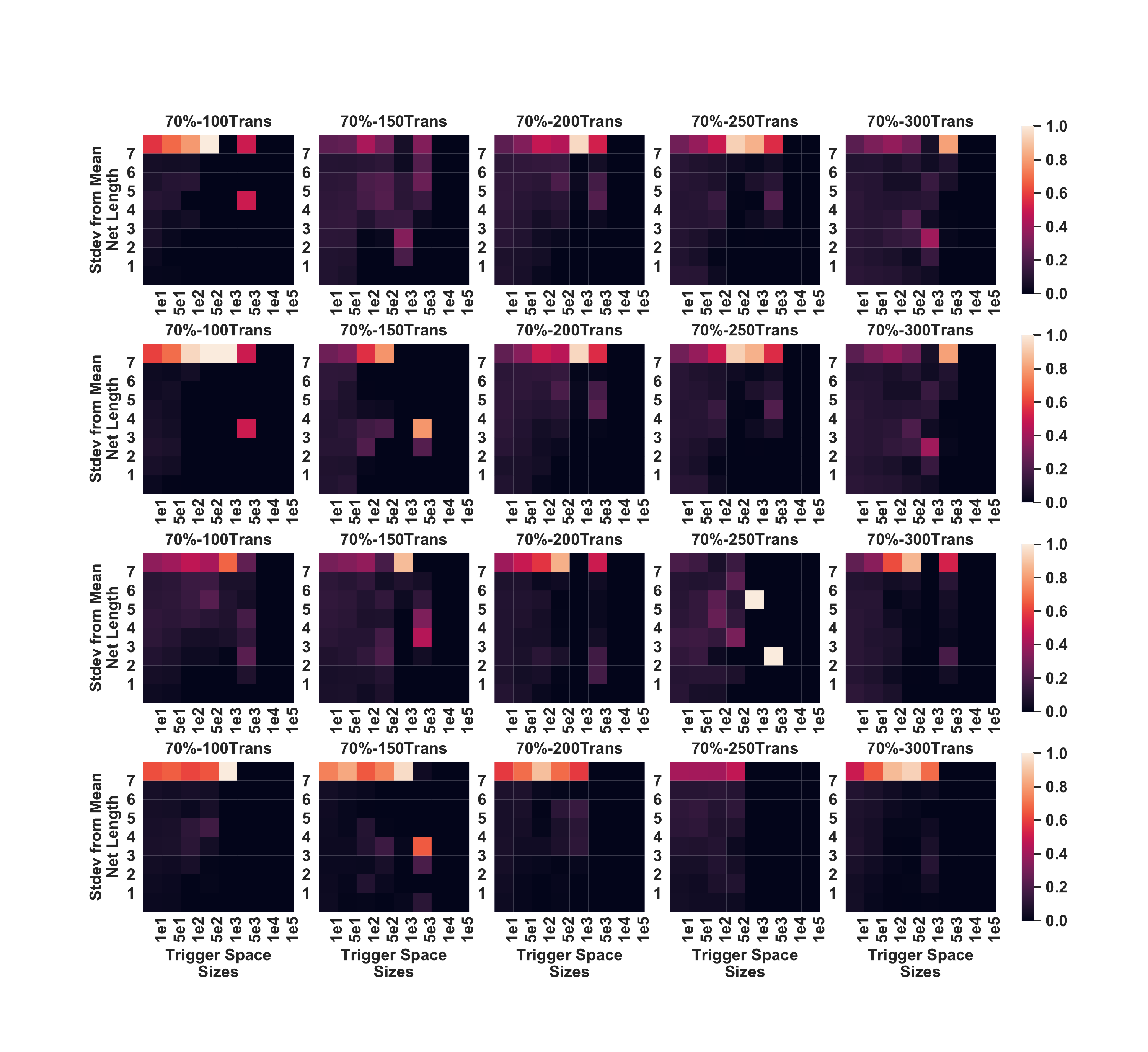}}
\vspace*{-0.08\textwidth}
\caption{\footnotesize \textit{Route Distance Metric for OR1200 at 70\% Density.} Same as Fig.~\ref{fig:all_rd_50density}, except a target density of 70\% was held across each layout.}
\label{fig:all_rd_70density}
\end{figure*}

\begin{figure*}
\centering
\makebox[\textwidth][c]{\includegraphics[width=1.1\textwidth]{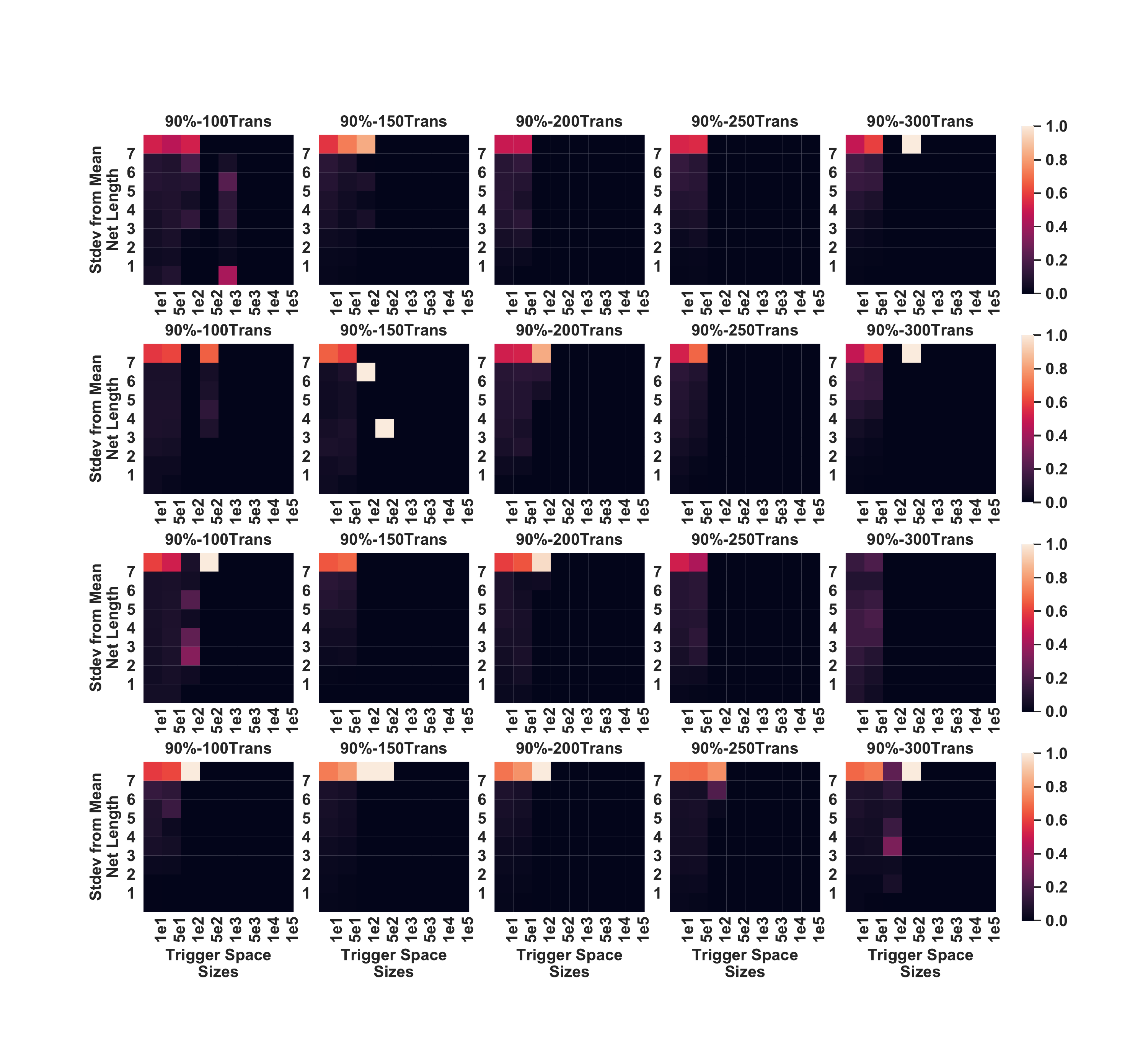}}
\vspace*{-0.08\textwidth}
\caption{\footnotesize \textit{Route Distance Metric for OR1200 at 90\% Density.} Same as Fig.~\ref{fig:all_rd_50density}, except a target density of 90\% was held across each layout.}
\label{fig:all_rd_90density}
\end{figure*}